\documentclass{pasa}%

\usepackage{graphicx}
\usepackage{amsmath}
\usepackage{gensymb} 
\usepackage{xfrac}
\usepackage[flushleft]{threeparttable}

\usepackage{aas_macros}
\usepackage{hyperref} 
\hypersetup{colorlinks,citecolor=blue,linkcolor=blue,urlcolor=blue}

\newcommand{\rednew}[1] {#1}
\newcommand{\red}[1] {#1}

\newcommand{\sensfluence}{300} 
\newcommand{\nfrbmerit}{77} 

\newcommand{\ntestpulsars }{11}

\title[A commensal FRB search pipeline for the MWA]{A commensal Fast Radio Burst search pipeline for the Murchison Widefield Array}

\author[M.~Sokolowski et al.]
{M.~Sokolowski$^{1}$\thanks{marcin.sokolowski@curtin.edu.au}, 
I. S. Morrison$^{2,1}$, 
D. Price$^{1}$, 
G. Sleap$^{1}$, 
B. Crosse$^{1}$, 
A. Williams$^{1}$, 
L. Williams$^{1}$, 
C. James$^{1}$,
B.~W.~Meyers$^{1}$,
S. McSweeney$^{1}$,
 N. D. R. Bhat$^{1}$,
G. Anderson$^{1}$
\affil{$^1$International Centre for Radio Astronomy Research, Curtin University, Bentley, WA 6102, Australia}%
\affil{$^2$Australia Telescope National Facility, CSIRO Space and Astronomy, PO Box 1130, Bentley WA 6102, Australia}
}%

\jid{PASA}
\doi{10.1017/pas.\the\year.xxx}
\jyear{\the\year}

\hypersetup{draft}

\begin{document}

\begin{frontmatter}
\maketitle

\begin{abstract}
We present a demonstration version of a commensal pipeline for Fast Radio Burst (FRB) searches using a real-time incoherent beam from the Murchison Widefield Array (MWA). The main science target of the pipeline are bright nearby FRBs from the local Universe (including Galactic FRBs like from SGR 1935+2154) which are the best candidates to probe FRB progenitors and understand physical mechanisms powering these extremely energetic events. Recent FRB detections by LOFAR (down to 110\,MHz), the Green Bank Telescope (at 350\,MHz), and CHIME detections extending down to 400\,MHz, prove that there is a population of FRBs that can be detected below 350\,MHz. The new MWA beamformer, known as the `MWAX multibeam beamformer', can form multiple incoherent and coherent beams (with different parameters) commensally to any on-going MWA observations. One of the beams is currently used for FRB searches (tested in 10 kHz frequency resolution and time resolutions between 0.1 and 100\,ms). A second beam (in 1\,Hz and 1 s frequency and time resolutions respectively) is used for the Search for Extraterrestrial Intelligence (SETI) project. This paper focuses on the FRB search pipeline and its verification on selected known bright pulsars. The pipeline uses the FREDDA implementation of the Fast Dispersion Measure Transform algorithm (FDMT) for single pulse searches. Initially, it was tested during standard MWA observations, and more recently using dedicated observations of a sample of \ntestpulsars\, bright pulsars. The pulsar PSR J0835-4510 (Vela) has been routinely used as the primary probe of the data quality because its folded profile was always detected in the frequency band 200 -- 230\,MHz with typical signal-to-noise ratio $>$10, which agrees with the expectations. Similarly, the low dispersion measure pulsar PSR B0950+08 was always detected in folded profile in the frequency band 140 -- 170\,MHz, and so far has been the only object for which single pulses were detected. We present the estimated sensitivity of the search in the currently limited observing bandwidth of a single MWA coarse channel (1.28\,MHz) and for the upgraded, future system with 12.8\,MHz (10 channels) of bandwidth. Based on expected sensitivity and existing FRB rate measurements, we project an FRB detection rate between a few and a few tens per year with large uncertainty due to unknown FRB rates at low frequencies.
\end{abstract}

\end{frontmatter}

\section{INTRODUCTION}
Fast Radio Bursts (FRBs) are a recently-discovered class of astrophysical transients of predominantly extragalactic origin. They are highly energetic bursts at radio wavelengths, lasting only a few milliseconds and detectable from the distant Universe \red{(up to and perhaps beyond redshift z$=$1, e.g. FRB 20220610A with z=$1.016 \pm 0.002$ reported by \citet{doi:10.1126/science.adf2678})}, and as such have emerged as a frontier field of modern astrophysics \citep[reviews ][]{2022A&ARv..30....2P,2021Univ....8....9P,2019ARA&A..57..417C}. In just over 15 years, the number of FRBs ``sky-rocketed'' from a single Lorimer Burst \citep{2007Sci...318..777L} through a few tens of detections with Parkes (Murriyang) radio-telescope \citep{2013Sci...341...53T} to several hundreds \citep{2021ApJS..257...59C}. \red{The interferometric localisations and associations with host galaxies have enabled redshift measurements and ultimately confirmed the extra-galactic origin of FRBs \citep{2017Natur.541...58C,2017ApJ...834L...7T,2019Natur.572..352R,2019Sci...365..565B,2019Sci...366..231P,2020Natur.577..190M}. Furthermore, the interferometric localisations of several FRBs by the Commensal Realtime ASKAP Fast Transients (CRAFT) survey \citep{Macquartetal2010} on the Australian Square Kilometre Array Pathfinder (ASKAP) at 1.4\,GHz also enabled measurements} of the electron content of the Universe, and established the Macquart relation between the dispersion measure (DM) and redshift  \citep{2020Natur.581..391M}. With the increasing number of localised detections from ASKAP CRAFT, Deep Synoptic Array \citep[][DSA-110\footnote{\url{https://deepsynoptic.org}};]{2023ApJ...949L...3R} and other instruments, the precision of these measurements and significance of FRBs as cosmological probes will continue to  increase. In addition to using FRBs as cosmological tools, there have been on-going efforts to understand their progenitors and underlying physical mechanisms. In the early days of FRB research, Arecibo telescope discovered the first repeating FRB 121102 \citep{2014ApJ...790..101S}, \red{which led to the hypothesis that there are two distinct populations of FRBs: namely, repeating, and one-off.}

In the first few years, the FRB field was dominated by dish telescopes operating at GHz frequencies. Although the initial efforts at sub-GHz frequencies were unsuccessful, eventually FRBs were detected at 800\,MHz \citep{2017MNRAS.468.3746C} by the UTMOST telescope \citep{2017PASA...34...45B}. In 2018, Canadian Hydrogen Intensity Mapping Experiment (CHIME) came on-line, started to detect many FRBs, and became a true northern hemisphere ``FRB factory''. In 2021, CHIME published a catalogue of 536 one-off and 18 repeating FRBs \citep{2021ApJS..257...59C,2020Natur.582..351C} at 400 -- 800\, MHz, and more recently confirmed another 25 repeating FRBs \citep{2023ApJ...947...83A}. Their large sample of FRBs enabled statistical and morphological studies of the FRB population \citep{2021ApJ...923....1P}. 
These results indicate that physical properties of one-off and repeating FRBs are different, which suggests different underlying populations of sources or differences in the local environments of the two classes. The main limitation of the CHIME telescope has been the localisation accuracy, though the upcoming \red{outrigger} project will provide sub-arcsecond localisation precision \citep{2023arXiv230410534S} and guarantee that CHIME will also contribute significantly more to cosmological studies. Intriguingly, many CHIME FRBs detected down to 400\,MHz appear not to be scattered (modulo CHIME's limitations to measure scattering), which suggests that many of the CHIME FRBs should also be detectable at frequencies below 400\,MHz.

\begin{table*}
\caption{Summary of past, present and future non-targeted wide-field and all-sky searches for low-frequency FRBs. Only \citet{2020ApJ...904...92P} (1$^{st}$ line) detected one FRB.}
\vspace{-0.3cm}
\centering
\begin{tabular}{@{}ccccccccc@{}}
\hline
\textbf{Reference} & \textbf{T}$^a$ & \textbf{Frequency} & \textbf{Detection} & \textbf{Time} & \textbf{Band-} & \textbf{FoV} & \textbf{Obs.} & \textbf{Figure$^b$}   \\
                   & & \textbf{range}     & \textbf{threshold}    & \textbf{resolution}  &  \textbf{width}    &  \textbf{[deg}$^2$\textbf{]} & \textbf{Time} & \textbf{of merit} \\
                   & & \textbf{[MHz]}     & \textbf{[Jy\,ms]}  &  \textbf{[ms]} &  \textbf{[MHz]}    &  & \textbf{[days]} & $\propto \text{N}_\text{FRB}$\\ 
\hline
\citet{2020ApJ...904...92P} & G & 350 & 1.26 & 0.08192 & 100 & 0.27 & 173.6 & 2.1 \\ 
\citet{2020MNRAS.493.4418R} & J & 332 & 46 & 0.256 & 64 & 0.61 & 58 & 0.62 \\ %

\citet{Coenen_et_al_2014} & L & 140 & 71 & 0.66 & 6  & 75 & 9.7 & 1.09 \\ 

\citet{2015MNRAS.452.1254K} & L & 145        & 310    & 5     & 6     & 24 & 60.25 & 1.6     \\ 
\citet{2015AJ....150..199T} & M & 139 -- 170 & 700    &  2000 & 30.72 & 610 & 0.44 & 3.6 \\ 
\citet{2016MNRAS.458.3506R} & M & 170 -- 200 & 223500 & 28000 & 30.72 & 452 & 3.3 & $2.4 \times 10^{-7}$     \\ 
\textbf{MWA IC (this work)} & \textbf{M} & \textbf{210 -- 223$^c$} & \textbf{400}$^d$ & \textbf{10} & \textbf{12.8} & \textbf{400} & \textbf{365} & \textbf{53} \\
\citet{2022aapr.confE...1S}  & C & 50 -- 350 & \sensfluence$^d$ & 10 & 40    & 12000$^e$ & 365 & \nfrbmerit \\ 

\hline
\end{tabular}
\begin{flushleft}
\begin{tablenotes}
\small
\item \textbf{$^a$} Telescopes: G - GBT (100\,m), J - Lovell telescope at Jodrell Bank (76\,m), L - LOFAR, M - MWA ($\sim$60\,m), C - all-sky FRB monitor CHASM implemented on SKA-Low stations (effective size 34\,m at 100\,MHz and 20\,m at 200\,MHz) described by \citet{2022aapr.confE...1S}. The values in brackets are dish diameters or equivalent for the aperture arrays
\item $^b$ Figure of merit  M = FoV $\times$ $\text{F}^{-3/2} \times$  $\text{T}_{obs} / \delta t$ (equation~\ref{eq_fom}) as defined in \red{\citep{2004NewAR..48.1459C,2008ASPC..395..225C,2009ASPC..407..318H,Macquartetal2010}. Here, M was divided by $\text{M}_0$}, where $\text{T}_{obs}$ is the observing time, and $\delta t$ is the time resolution. $M_0$ is M calculated for survey parameters of the survey by \citet{2020ApJ...904...92P}, which detected 1 FRB in about 174 days. \red{This figure of merit increases with the increasing FoV, sensitivity, total observing time, and also with the improved time resolution ($\delta t$). Instead of detecting a single pulse during observing time $T_{obs}$, the higher time resolution enables detection of multiple ($\sim T_{obs} / \delta t$) short pulses ($\le \delta t$) leading to more FRB detections.}
\item $^c$ The exact frequency range is 210.56 to 223.36\,MHz
\item $^d$ 10$\sigma$ threshold
\item $^e$ Above elevation 25\degree 

\end{tablenotes}
\end{flushleft}
\label{tab_low_freq_searches}
\end{table*}

\subsection{FRB searches at frequencies below 350\,MHz}
\label{subsec_frb_searches_below_350MHz}

Despite the success of CHIME at frequencies above 400\,MHz, and ongoing efforts at lower frequencies, there have only been a few FRB detections at frequencies below 400\,MHz. The initial searches by LOFAR \citep{Coenen_et_al_2014, 2015MNRAS.452.1254K} did not detect any FRBs. Similarly, efforts using the  Murchison Widefield Array \citep[MWA;][]{2013PASA...30....7T,2018PASA...35...33W}, failed mainly because of the limited on-sky time and signal processing constraints (time and frequency resolutions $\ge$0.5\,s and 1.28\,MHz respectively) that limited sensitivity to short pulses to $\gtrsim$500\,Jy\,ms \citep{2016MNRAS.458.3506R,2016Natur.530..453K,2015AJ....150..199T}. Table~\ref{tab_low_freq_searches} summarises these earlier \rednew{non-targeted} FRB searches at low frequencies.

Between 2017 and 2019, \citet{2018ApJ...867L..12S} conducted an MWA campaign and co-observed (shadowed) the ASKAP field of view (FoV). During these observations ASKAP detected several FRBs, and two of them during favourable nighttime. Unfortunately, calibration of daytime MWA data was very difficult until an observing strategy placing the Sun in the null of the primary beam was implemented in late 2019 \citep{2019PASA...36...46H}. However, it turned out that neither of these two ASKAP FRBs was simultaneously (after correcting for dispersion delay) detected in the 0.5-sec/1.28\,MHz images from the MWA. The upper limit on flux density of these FRBs at 200\,MHz led to constraints on the properties of the immediate surroundings of the FRB sources (e.g. on the size of the absorbing region) demonstrating the potential applications of the low-frequency observations (including non-detections). More recently, \citet{2023MNRAS.518.4278T} used archival MWA high-time resolution data from the Voltage Capture System \citep[VCS;][]{2015PASA...32....5T} to look for pulses from a modest sample of FRBs (one ASKAP and four CHIME). Although they did not detect any pulses from these FRBs, similar targeted searches with the MWA and other low-frequency telescopes have significant potential to detect low-frequency pulses from repeating FRBs. 

This was the case of one of the CHIME repeating FRBs 20180916B with a regular (hence predictable) activity period, which was detected by LOFAR \citep{2021ApJ...911L...3P,2021Natur.596..505P} at frequencies down to even 110\,MHz -- the first ever FRB detection below 300\,MHz. Commensal observations of FRB 20180916B with CHIME, LOFAR and Apertif reported by \cite{2017ursi.confE...2M} revealed that low frequency emission was usually not detected when high frequency emission was and vice-versa, which is a possible explanation to earlier MWA non-detections of ASKAP FRBs by \citet{2018ApJ...867L..12S}. This so-called chromaticity window further supports the need to conduct independent searches for low-frequency FRBs, as the low-frequency signals may not be simultaneous with bursts at higher frequencies. Additionally, a one-off FRB 20200125A was discovered by the Green Bank Telescope (GBT) at 350\,MHz \citep{2020ApJ...904...92P}. These detections, together with CHIME FRBs extending down to 400\,MHz, ultimately prove that FRBs can be detected at low radio frequencies.

\subsection{FRB progenitors and models}
\label{sec_frb_theory}

\red{Although the field has achieved} significant progress on both observational and theoretical fronts (see \citet{2022A&ARv..30....2P} for the recent review), physical mechanisms and FRB progenitors remain unexplained. A detailed summary of FRB models exceeds the scope of this paper but very good reviews of existing theoretical models can be found in Section 9 of \citet{2019A&ARv..27....4P} or in the FRB Theory Catalogue\footnote{\url{https://frbtheorycat.org/index.php}} \citep{2019PhR...821....1P}. 
\rednew{In short, the leading models for repeating FRBs relate them to magnetars \citep{2017ApJ...841...14M,2018MNRAS.481.2407M}, which are highly magnetised ($\sim 10^{15}$\,G) neutron stars \citep[e.g.][]{2018Ap&SS.363..242L}. Such long-lived stable magnetars can produce coherent radio pulses in a similar way to pulsars (dipole radiation) and be observed as repeating FRBs during their activity periods.}

\rednew{On the other hand, one-off FRBs are hypothesised to be produced in cataclysmic events, such as a collapse of a super-massive neutron star (NS) \citep[e.g.][]{2014A&A...562A.137F} as its rotation slows down due to magnetic braking. A super-massive neutron star can be formed in a cataclysmic event like an NS-NS merger \citep{2013PASJ...65L..12T,2016MNRAS.459..121C,2014ApJ...780L..21Z,2017LRR....20....3M} leading to a super-massive short-lived (seconds to hours) magnetar, which emits coherent radio bursts during its short lifetime and ultimately collapses to a black hole \citep{2019MNRAS.489.3316R,2023arXiv231204237R}.} 

\rednew{Hence, magnetars are one of the main contenders for progenitors of both repeating and at least some non-repeating FRB. }The magnetar model is strongly supported by the detections of FRB-like $\sim$MJy radio pulses from the Galactic Soft Gamma Repeater SGR 1935+2154 \citep{2020Natur.587...59B,2020Natur.587...54C}, which was the only FRB-like event observed at other electromagnetic wavelengths. On the other hand, the more recent detection of FRB 20200120E \citep{2022Natur.602..585K} pinpointed to a globular cluster (GC) slightly challenges the magnetar model as this kind of young neutron star is not expected to be present in GCs. An alternative model for FRBs is that they are due to extremely bright and short (even ns duration) \red{pulses similar to so-called supergiant pulses emitted by pulsars like PSR B0531+21 (aka Crab) \citep{2016MNRAS.457..232C,2016MNRAS.458L..89C}}. 

\rednew{Although there is a general consensus that FRBs are produced by a coherent emission process, the exact radiation mechanisms are yet to be determined. In pulsar-like models, FRBs are produced by coherent emission processes occurring in the magnetosphere close ($\lesssim 10^4$\,km) to the surface of neutron star via magnetic reconnection\citep[e.g.][]{2021ApJ...922..166L} or curvature radiation \citep{2017MNRAS.468.2726K}. On the other hand, in GRB-like models, coherent radio pulses are generated further away from the surface ($\gtrsim 10^5$\,km) of neutron star via synchrotron maser mechanism in the forward shock of the flare of material ejected from a magnetar as it collides with the surrounding medium \citep{2019MNRAS.485.4091M}. Comprehensive discussions can be found in the recent reviews by \citet{2022A&ARv..30....2P} and \citet{2021Univ....8....9P}.}

\rednew{The same physical processes can also produce low-frequency radio signals ($\lesssim$300\,MHz). However, radio signals at these frequencies may be suppressed by several mechanisms. At frequencies below plasma frequency $\lesssim$90\,MHz they are quenched by plasma absorption, while at frequencies 90\,MHz$\,\lesssim \nu \lesssim$\,300\,MHz by free-free absorption in the NS's dense immediate surroundings or ejected material \citep{2021Univ....8....9P}. Therefore, detection of low-frequency FRBs may be possible only in low density environments where absorption is negligible, which may be the case at least in some progenitor systems, like FRB 190816B~\citep{2021ApJ...911L...3P} and 200125A~\citep{2020ApJ...904...92P}.}

\rednew{FRB 180916B was detected in a targeted LOFAR search for low-frequency pulses from a CHIME repeating FRB with a known 16-day periodicity. As discussed earlier, repeating FRBs are believed to be due to stable magnetars, while their periodicity may be caused by interactions with the stellar wind from a companion star in the binary system with a massive/neutron star~\citep{2020ApJ...893L..26I,2020ApJ...893L..39L}  or precession of the magnetar's spin axis~\citep{2020ApJ...892L..15Z,2020RAA....20..142T}. Both models predict frequency dependent activity window and other characteristics which can be tested by simultanous observations at high and low frequencies~\citep{2021Natur.596..505P}.}

\rednew{Finally, the most promising physical scenario leading to one-off low-frequency FRBs are events associated with short Gamma-Ray Bursts (SGRBs), which are also linked to NS-NS mergers. SGRBs seem to occur in low density environments \citep{2015ApJ...815..102F}. Hence, low-frequency radio signals produced at various stages of NS-NS merger can avoid absorption and be detected by low-frequency radio-telescopes \citep{2019MNRAS.489.3316R}. A potential association of a coherent radio pulse with short GRB 201006A was recently reported by \citet{2023arXiv231204237R}.}


\subsection{A hunt for bright, nearby FRBs}
\label{sec_hunt_for_bright}

Similarly to other astrophysical phenomena (e.g. Gamma-Ray Bursts (GRBs)) multi-wavelength observations may hold the key to explaining the FRB enigma.  However, except the special case of the Galactic FRB from SGR 1935+2154, so far no FRB has been detected at other electromagnetic wavelengths than radio. Detection of more Galactic FRB-like events linked to magnetars, young NSs or other objects will provide essential observational evidence to support or disfavour theoretical models of FRBs. 

The best candidates for the first multi-wavelength detections are bright FRBs from the local Universe. Therefore, nearby FRBs are of great interest for detailed studies of FRB host galaxies, progenitors and local environments. Accurate localisations of such nearby FRBs may pinpoint their host galaxies and even specific objects within host galaxies (e.g.~\cite{2022Natur.602..585K}) which will uncover information about their progenitors. Fast and precise localisation of bright nearby FRBs can lead to detections over a broad range of the electromagnetic spectrum (optical, gamma, X-rays etc) and/or in other messengers such as gravitational waves (GWs), which will be particularly useful for explaining the underlying physics. \cite{2022MNRAS.509.4775J} provide evidence that many FRBs may originate from nearer in the Universe than their DMs suggest. Detections and localisations of nearby FRBs from the Local Group, Virgo Cluster etc. can confirm these findings and verify these observation models. 

Identifying links between FRBs and other transient events such as GRBs, GW events, or binary neutron star (BNS) mergers will help to understand all these processes and develop a unified model. The MWA automatic response system \citep{2019PASA...36...46H} enabled searches for coherent radio emission from short and long GRBs \citep{2021PASA...38...26A,2022PASA...39....3T,2022MNRAS.514.2756T}. Although so far unsuccessful, they may eventually lead to positive detections as the capabilities and sensitivity of the MWA improve. Similarly, a detection of an FRB accompanying GWs from nearby ($\sim$40\,Mpc) BNS mergers like \citet{2017ApJ...848L..12A} would confirm the link between these two classes of events suggested by theories \citep{2019MNRAS.489.3316R,2016MNRAS.459..121C,2013PASJ...65L..12T}. The MWA is particularly well suited to detect potential FRB-like counterparts of GW events as described in \cite{2019MNRAS.489L..75J}, and supported by the recent associations of the CHIME FRB 190425A with GW190425~\citep{2023NatAs...7..579M,2023MNRAS.519.2235P}.  
Furthermore, as described by \citet{2023PASA...40...50T}, the MWA is also in a perfect geographical location to maximise the chances of detecting FRB counterparts of GW events detected by the LIGO-Virgo-KAGRA \citep[LVK;][]{2018LRR....21....3A}. 

Such bright FRBs can potentially be detected in the MWA incoherent beam, which can trigger the recording of high time resolution complex voltages leading to the required accurate localisations. The MWA is currently the only low frequency (70 -- 300\,MHz) radio telescope in the southern hemisphere, and therefore it is important to increase and take full advantage of its capabilities for FRB and other high-time resolution science. In this paper we describe the initial version of the MWA real-time pipeline for FRB searches in the incoherent beam (IC).

This paper is organised as follows. In Section~\ref{sec_mwa} we describe the MWA telescope and the processing pipeline forming real-time incoherent beams. In Section~\ref{sec_realtime_frb_pipeline} we present the real-time FRB search pipeline using the incoherent beam. We also discuss sensitivity predictions for pulsars and FRBs with the pipeline using a single (1.28\,MHz) and ten (12.8\,MHz) coarse frequency channels. In Section~\ref{sec_verification} we present results of the pipeline verification using dedicated observations of selected bright pulsars. Finally, in Section~\ref{sec_summary_future} we summarise and discuss future work.

\section{MWA Telescope}
\label{sec_mwa}

The Murchison Widefield Array \citep[MWA;][]{2013PASA...30....7T,2018PASA...35...33W} is a precursor of the low-frequency Square Kilometre Array telescope \cite[SKA-Low;][]{2009IEEEP..97.1482D}\footnote{www.skatelescope.org}. It is located in the  Murchison Radio-astronomy Observatory (MRO) in a Radio Quiet Zone \citep[RQZ;][]{7731554} in Western Australia, which is a highly desirable location for high sensitivity FRB searches. Originally designed as an imaging instrument, at an early stage the MWA was converted into a multi-purpose telescope capable of recording high-time resolution voltages suitable for pulsar and FRB science. Initially, it was composed of \red{$N_{ant}=$\,128} small aperture arrays (``tiles'') consisting of 16 bow-tie dipole antennas arranged in a $4 \times 4$ array. \red{The individual antennas in a tile are analogue beamformed, hence, each tile performs as a single antenna unit (i.e. small low-frequency ``dish''). The maximum baseline between the tiles was originally approximately 3\,km}. In 2018, the MWA was upgraded \citep{2018PASA...35...33W} and extended with additional 128 tiles. The compact configuration (maximum baseline $\approx$740\,m), targeting mainly Epoch of Reionisation and pulsar science, comprises 72 tiles arranged in two hexagonal grids (``the hexes'') of 36, and 56 tiles from the innermost region of the original array. The hexes provide redundant baselines enabling a redundant calibration scheme, improving sensitivity of power spectrum measurements, while the larger synthesised beam enables computationally affordable pulsar searches \citep{2023PASA...40...21B,2023PASA...40...20B}. On the other hand, the extended configuration, including 56 long-baseline tiles with the maximum baseline $\approx$5.3\,km, improves imaging spatial resolution by nearly a factor of two and reduces classical and confusion noise. 
\red{The signals from individual 16 antennas within each tile are summed in analogue beamformers. Hence, in standard observing modes the information about signals from individual dipole antennas are not preserved and an MWA tile performs as an individual antenna unit of the MWA telescope. Therefore, in this paper the variable $N_{ant}=$\,128 (or currently 144) is the number of the used MWA tiles, and it does not refer to individual MWA dipoles.}

The MWA receivers channelise the full 70-300 MHz received bandwidth into 1.28 MHz wide coarse channels. The MWA can process 30.72 MHz of instantaneous bandwidth by selecting an arbitrary subset of 24 coarse channels. \red{These selected channels can be arranged in a continuous block of 24 (30.72\,MHz of continuous bandwidth) or be an arbitrary selection of 24 channels (the so called ``picket-fence'' mode).}

The original receivers and correlator \citep{2015PASA...32....6O} enabled operation of 128 tiles at any given time. Therefore, for the last 5 years the MWA has been operating in either compact or extended configuration with a different set of tiles connected to 16 receivers. However, the recent commissioning of the new MWAX correlator \citep{2023PASA...40...19M} opens a possibility of increasing the number of tiles to 256 once additional receivers are commissioned and deployed at the MRO. Recently two new receivers have been commissioned (18 in total), and the MWA is currently operating at 144 tiles.

\subsection{Real-time incoherent beam}

The new MWAX correlator also provides new beamforming capability, which can form multiple real-time \red{tied-array} (coherent) and incoherent beams \red{at the frequency of an on-going MWA observation (commensality of the pipeline). Thus, the pipeline forms the beams using selected (currently 1 out of 24) coarse channels of an on-going MWA observation.} These beams can be formed in real-time and their number is limited only by the available compute hardware. The observing bandwidth is also limited by the compute hardware and the throughput of the network connection between the MRO and the computing centre on the Curtin University campus (Curtin) as UDP packets are currently transmitted from the MRO to Perth (where beamforming is performed) over a 100\,Gbit link. This link is also used for archiving standard MWA observations. Hence, a maximum of about 10 channels (12.8\,MHz) can be transmitted to ensure that the bandwidth of the connection is not saturated and MWA operations are not affected. In the future, as the number of MWA tiles increases (ultimately to 256) and so do the bandwidth requirements of the standard MWA observations, the system may be deployed at the MRO in order to be independent of the limitations of the Perth -- Curtin link. 

Once UDP packets are captured the signals from each tile are fine channelised and then the signal powers of each tile (within each channel) are incoherently summed to form a channelised incoherent beam:

\begin{equation}
I_c(t) = \sum_{a=0}^{N_{ant}} w_c^a I_c^a(t),
\end{equation}

where $I_c(t)$ is the incoherent sum in channel c at time t, $I_c^a(t)$ is the power from antenna a in channel c at the time t, \red{$N_{ant}$ is the number (128 or 144) of used MWA tiles (each tile performs as an individual antenna unit of the MWA telescope), and  $w_c^a$ is the weight of antenna $a$ at frequency channel $c$. These weights are currently set to 1 but in future versions can be set to zero in order to flag (exclude) broken antennas (or RFI affected channels) from the incoherent sum. Weights can also be used for sub-arraying by setting the weights of unused antennas to zero, or some other value in the range [0,1] to apply a specific weighting schema across the array.} The summed powers are optionally averaged in time as requested by the parameters specified in a beamformer configuration file: 

\begin{equation}
\overline{I_c(t)} = \frac{1}{K} \sum_{k=0}^{K} I_c(t + k\Delta t),
\end{equation}

where $\Delta t \approx$0.78 microseconds is the sample period, $\Delta T$ is the requested time resolution specified in the configuration file, and $K=\Delta T/\Delta t$ is the number of time samples in the requested averaging time bin. \red{The incoherent beam preserves the large MWA FoV ($\sim$20$\times$20 deg$^2$ at 200\,MHz) at the expense of lower sensitivity (as discussed in Section~\ref{sec_expected_sensitivity}). It is also computationally more tractable and suitable for real-time searches in comparison to tied-array beamforming \citep{2019PASA...36...30O,2020PASA...37...34M,2022PASA...39...20S}, which has higher sensitivity but requires more compute power to tessellate the entire FoV with narrow beams (the approximate half-power beamwidth is $\lambda/B$ radians, where $\lambda$ is the observing wavelength and B maximum distance between two MWA tiles). }
Multiple incoherent beams with different channelisation and time averaging can be formed simultaneously.  The system is fully commensal, and incoherent beams are formed from complex voltages generated during all standard (correlator and voltage capture mode) MWA observations. 

The current pilot system forms only 3 incoherent beams 
\red{using a single coarse channel (1.28 MHz) selected from the 24 coarse channels of the ongoing MWA observation. The small observing bandwidth of the current system (1.28 MHz)} limits the sensitivity of the FRB search by a factor of $\approx$3 in comparison to the future search using 10 coarse channels. The three beams are currently generated for: (i) FRB search (typically 1 to 100\,ms time resolution and 10 kHz frequency resolution), (ii) Search for Extra-Terrestrial intelligence (SETI) in 1 s and 1\,Hz time and frequency resolutions respectively, and (iii) real-time folding with a specified period to verify detection of a test pulsar that is in the MWA FoV of an observation. In this latter case, no channelisation is performed, i.e. the time resolution is the coarse channel sample period of $\approx$0.78 microseconds and the frequency resolution is the full coarse channel width of 1.28 MHz. The planned future improvements in the pipeline, including increase of the observing bandwidth, are described in Section~\ref{sec_summary_future}. 

\begin{figure*}[t]
\begin{center}
\includegraphics[width=0.95\textwidth,angle=0]{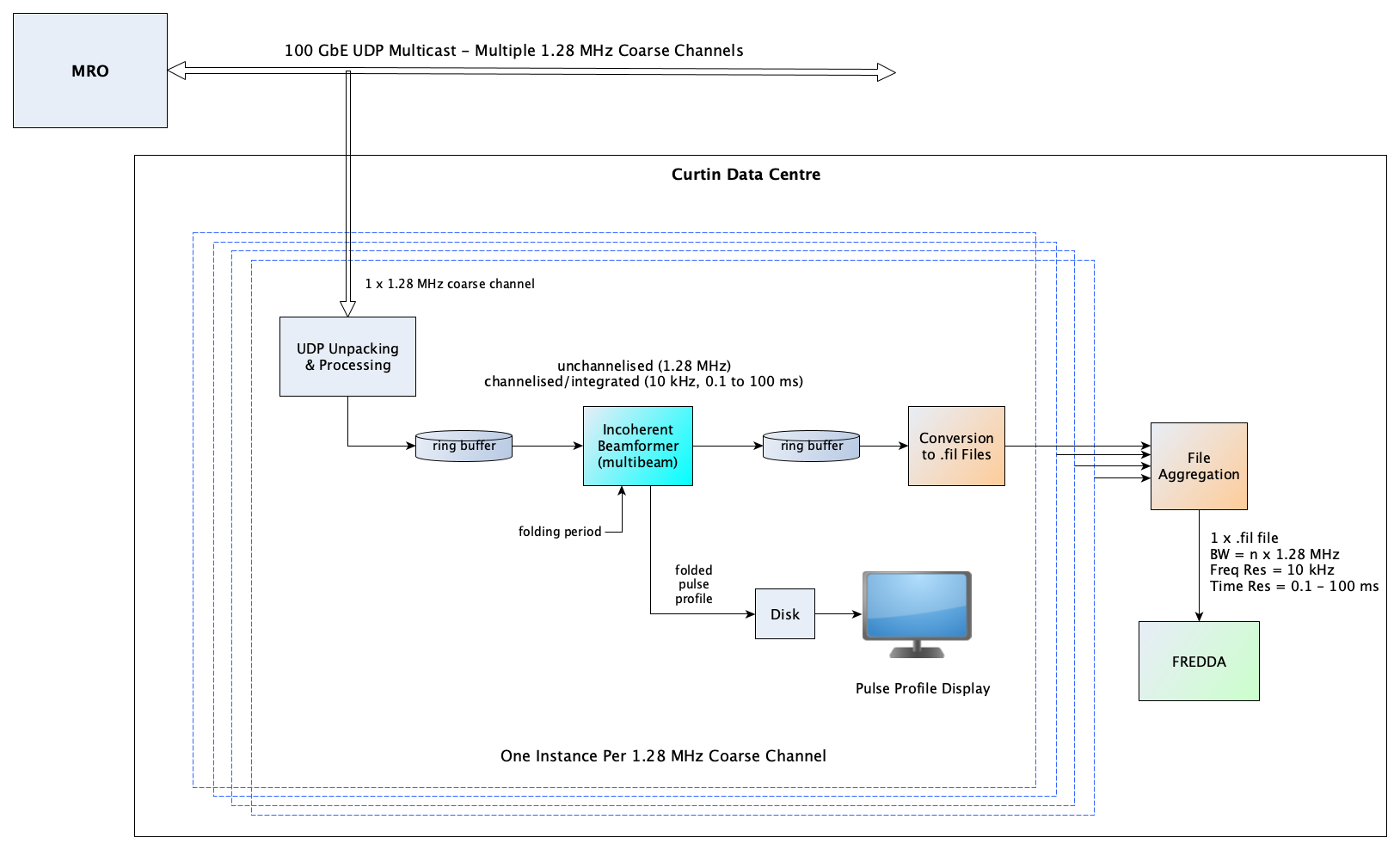} 
\caption{Block diagram of the MWA FRB search pipeline including the real-time folding feature, which can be used to verify detection of specified pulsars (within the MWA primary beam) and data quality in real-time.}
\label{fig_piepline_diagram}
\end{center}
\end{figure*}

\subsection{Hardware and software used for real-time beamforming}

This initial pilot pipeline runs on a single server (hosted in a server room on Curtin campus) with the following specifications:
\begin{itemize}
\item \verb|CPU|: Dual socket Intel Xeon E5-2620 running at 2.10GHz
\item \verb|Memory|: 512 GB
\item \verb|GPU|: 1 x NVIDIA RTX A4500 (20 GB RAM)
\item \verb|Storage|: 2 RAID 5 arrays of 11 x 4.5 TB disks resulting in two volumes of 46 TB formatted as xfs
\item \verb|Network|: 1 x Mellanox ConnectX-3 with a 40 Gbps fibre optic connection to a Cisco Nexus 9504 switch which provides the multi-cast UDP data from the MRO
\end{itemize}

This system is configured with a net booted Ubuntu 16.04 LTS operating system from a head node server (allowing more compute nodes to be added easily in the future). 

The software stack includes the following components:
\begin{itemize}
\item \verb|mwax_u2s|: This program captures a single coarse channel from the MWA multicast UDP datastream and organises the data into sub-observation files (known as ``subfiles''), each containing 8 seconds of observation data, written to a RAM disk (in this case the /dev/shm RAM disk filesystem). This is the same process that is run on the MWAX correlator servers at the MRO \citep{2023PASA...40...19M}.
\item \verb|mwax_mover|: This python process detects new subfiles created in the /dev/shm filesystem and loads the data into a \verb|PSRDADA| ring-buffer \citep{2021ascl.soft10003V}, whilst also appending beamformer configuration information, read from a configuration file, to the \verb|PSRDADA| ring-buffer header. The beamformer configuration information includes the number of incoherent beams to generate and each beam's frequency and time resolution.
\item \verb|mwax_db2multibeam2db|: This binary performs fine channelisation (using the \verb|cuFFT|\footnote{\url{https://developer.nvidia.com/cufft}} library) and then carries out the beamforming task based on parameters passed via the \verb|PSRDADA| ring-buffer header. The beamformed data, which might be one to many beams, are then written to an output \verb|PSRDADA| ring-buffer.
\item \verb|mwax_beamdb2fil|: This program reads the beamformed data from the output \verb|PSRDADA| ring-buffer and writes it to one of the 46TB RAID 5 volumes as a filterbank file.
\item \verb|process_new_fil_files_loop.sh|: This script detects new filterbank files, executes FREDDA and creates images with dynamic spectra of the resulting FRB candidates (Section~\ref{sec_realtime_frb_pipeline}). In a similar way new filterbank files will be processed to search for SETI (e.g. using TurboSETI software), which will be described in a separate publication \red{(Price et al., in preparation)}.
\end{itemize}

The diagram of the pipeline is shown in Figure~\ref{fig_piepline_diagram}. Since the multicast UDP data from the MRO is the exact same data that the MWAX correlator processes, we have been able to reuse some of the existing MWAX components (\verb|mwax_u2s| and \verb|mwax_mover|) and architecture (\verb|PSRDADA| ring-buffers) for this pipeline, which has reduced development and testing time/effort significantly.

The full software stack is deployed using the \verb|Ansible|\footnote{Ansible is an open source IT automation tool which allows scripting of software installations and configurations. See: /\url{https://www.ansible.com}} software tool, in order for operating system and software changes to be documented, repeatable, source controlled and easier to troubleshoot.

A constantly running monitor and control daemon allows remote monitoring, as well as the ability to remotely stop and start each process.

\section{Real-time FRB search in MWA IC beam}
\label{sec_realtime_frb_pipeline}

The resulting incoherent beams (sums) are saved as \textsc{filterbank} files. Separate files are formed for each MWA observation (typically of a few minute duration) and for incoherent beams with different parameters. These \textsc{filterbank} files are processed in real-time by FRB search software FREDDA \citep{2019ascl.soft06003B}. They can also be processed off-line using \rednew{standard pulsar software, such as PulsaR Exploration and Search TOolkit }\citep[PRESTO;][]{2011ascl.soft07017R}\rednew{. Off-line processing using PRESTO was performed on} observations of selected pulsars in order to confirm detection of their folded profiles. FREDDA saves the resulting FRB candidates to text files, which include basic information such as signal-to-noise (SNR), time, dispersion measure (DM) and \red{pulse width (in milliseconds)}, and can be used for further automatic analysis and/or visual inspection.

\subsection{Expected sensitivity of the FRB search using MWA incoherent beams}
\label{sec_expected_sensitivity}

\begin{figure*}[t]
\begin{center}
\includegraphics[width=0.95\textwidth,angle=0]{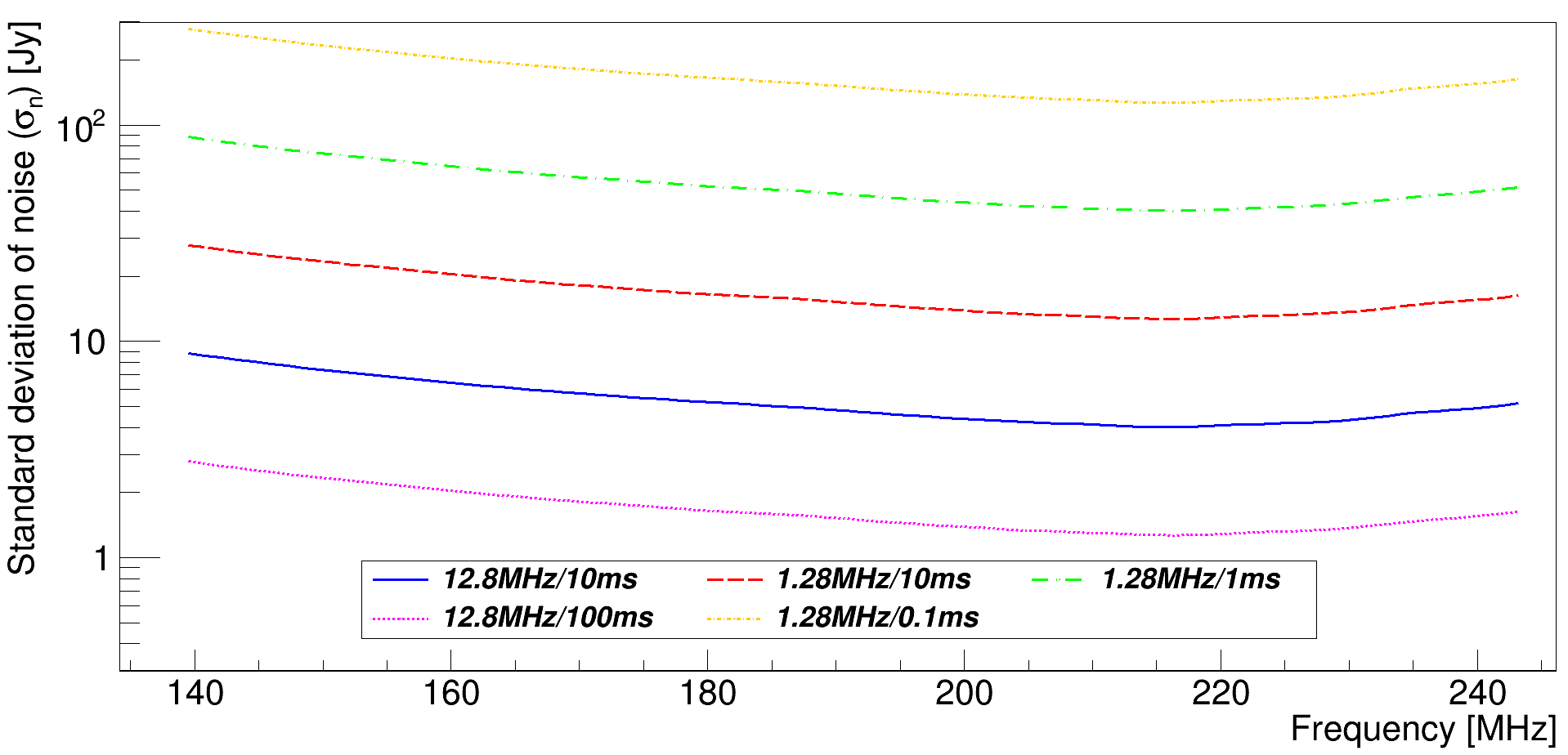}
\caption{Standard deviation of expected noise (sensitivity) as a function of frequency for a zenith-transiting source in the ``cold'' (i.e. low sky noise) part of the sky (RA=0\,h) using observing frequency bandwidth of 1.28 and 12.8 MHz (1 and 10 channel respectively) in 0.1, 1, 10\,ms, 100\,ms time resolutions. Note that some combinations, for example 10\,channels/10\,ms and 1\,channel/100\,ms, are equivalent due to the structure of the radiometer equation~\ref{eq_radiometer}. The best sensitivity (minima of the curves) is always at $\approx$216 MHz reaching approximately 127, 40, 12.7, 4 and 1.3 Jy for the curves 1.28MHz/0.1ms,  1.28MHz/1ms,  1.28MHz/10ms, 12.8MHz/10ms and 12.8MHz/100ms (from top to bottom) respectively.}
\label{fig_sensitivity}
\end{center}
\end{figure*}

The main advantage of the pipeline is that it can form incoherent (IC) sum and perform FRB and SETI searches \red{over the entire MWA FoV} commensally to any on-going MWA observations, without the need for dedicated observing time. On the other hand, the main disadvantage is that the sensitivity of the search in IC is lower than coherent searches using \red{tied-array} beam by a factor r=$\sqrt{N_{ant}}$, where $N_{ant}$ is the number of antennas (i.e. MWA tiles). Hence, in the current configuration of the MWA, with $N_{ant}$=128, r$\approx$11.3, i.e. sensitivity is reduced by approximately an order of magnitude with respect to searches using the \red{tied-array} beam. \red{Tied-array} beamforming and searches, however, are computationally very expensive \citep{2022PASA...39...20S} and cannot be realised in real time with the current hardware. 

Due to the sensitivity limitations, the real-time search in IC sum is mainly targeting the brightest, nearby FRBs which may be rare, and can only be detected with sufficiently long on-sky time provided by the commensality of the pipeline. The sensitivity of the IC searches was estimated using the MWA Full Embedded Element (FEE) beam model \citep{2017PASA...34...62S}, and the expected SNRs for the selected test pulsars are shown in Table~\ref{tab_bright_pulsars}. Most of the pulsar parameters were obtained from the pulsar catalogue \textsc{psrcat}\footnote{\url{https://www.atnf.csiro.au/research/pulsar/psrcat/}}\citep{2005AJ....129.1993M}. \red{If a pulsar was detected, its mean flux density (m) was measured using its folded profile (procedure described in Appendix A). Otherwise, mean flux density was obtained from \textsc{psrcat} or from \citet{2022PASA...39...42L}}. Pulse widths (w) at low frequencies may be significantly higher than those in \textsc{psrcat} due to scattering. Therefore, whenever available, they were estimated using the MWA pulsar census performed by \citet{2017PASA...34...70X}, and these estimates were used if the discrepancy was larger than 50\%. We used mean flux density, pulse width and pulsar period to estimate expected SNR of single pulse and peak in folded profile according to the following procedure:

\begin{itemize}
\item \textbf{Peak flux density} was calculated assuming a ``top hat'' pulse shape according to the following equation: $f_p = m P / w$, where P is the pulsar period. \red{This simple method was applied to all the pulsars except the Vela pulsar, which is significantly scattered at low frequencies. The scattering tail in Vela mean profile was accounted for when calculating its peak flux density by using the method described in Appendix A.} \\
\item \textbf{Standard deviation of the noise $\sigma_n$} (i.e. sensitivity) was calculated using the System Equivalent Flux Density (SEFD). SEFDs for X and Y polarisations were calculated using the MWA FEE beam model for the pointing direction to a specific pulsar, and they were combined into Stokes I $\text{SEFD}_I$ according to the following equation:

\begin{equation}
\text{SEFD}_I = 0.5 \sqrt{ \text{SEFD}_X^2 + \text{SEFD}_Y^2 },
\label{eq_sefd_i}
\end{equation}

which is strictly valid at zenith and on the cardinal axes aligned with the dipoles, while approximate to within acceptable 20\% at elevations $\ge$30\degree \, \citep{2021A&A...646A.143S,2022A&A...660A.134S}.
Finally, for the incoherent beam used in this work $\sigma_n$ was calculated as: 
\begin{equation}
\sigma_n = \frac{SEFD_I}{\sqrt{B \cdot \delta t \cdot  N_{ant}}},
\label{eq_radiometer}
\end{equation}

where B is the observing bandwidth (1.28 $10^6$\,Hz for a single channel), $\delta t$ is the time resolution (between 0.0001 and 0.1\,s), and $N_{ant}$ is the number of antennas/tiles (128 or 144 depending on the date of observations).\\

\item \textbf{The expected SNR} for single pulse detections SNR$_{s}$ can then be calculated as SNR$_{s} = f_p / \sigma_n$.\\

\item \textbf{To calculate the expected SNR of folded pulse profiles}, standard deviation of the noise $\sigma_n^{f}$ (where $f$ stands for folded profile) was calculated according to the same equation~\ref{eq_radiometer}, but in this case $\delta t$ was the amount of time contributing to a single time bin ($T/N_{bin}$) in a folded pulse profile. Hence:

\begin{equation}
\sigma_n^{f} = \frac{SEFD_I}{\sqrt{B \cdot T/N_{bin} \cdot  N_{ant}}},
\label{eq_radiometer_folded}
\end{equation}

where T is total duration of the observation (typically between 300\,s and 600\,s) and $N_{bin}$ is the number of phase bins in the folded profile (hence total time per phase bin $T/N_{bin}$). Consequently, the expected SNR of folded pulse profile was calculated as $\text{SNR}_{f} = f_p / \sigma_n^f$

\end{itemize}

The resulting sensitivities (in Jy) as a function of frequency for 1 and 10 frequency channels worth of bandwidth (1.28 and 12.8\,MHz respectively), and combination of other parameters (integration times and channel width) are shown in Figure~\ref{fig_sensitivity}. This figure shows that due to a combination of the frequency dependence of the sky noise and MWA beam, the optimal sensitivity is expected at frequency $\approx$216\,MHz. The values of optimal sensitivity (in terms of flux density and fluence) at 216\,MHz for different number of frequency channels and time resolutions are summarised in Table~\ref{tab_sensitivity_jy}. Assuming a typical pulse width (w) of an FRB $\sim$10\,ms and the same time resolution of the IC beam, the presented system with 10 channels (12.8\,MHz bandwidth) should be able to detect 40\,Jy pulses with SNR=10, which corresponds to an FRB with a fluence of $\approx$400\,Jy ms. It is clear that the presented system will be able to detect only the brightest FRBs, exceeding fluences $\sim$200\,Jy\,ms, which are very rare. For example, Australian Square Kilometre Pathfinder (ASKAP), detected only one (FRB 180110) with fluence $\approx$420\,Jy\,ms \citet{2018Natur.562..386S}. Nevertheless, continuous observations can also lead to detections of $\sim$MJy\,ms pulses as those detected from SGR 1935+2154 in 2020 \citep{2020Natur.587...54C} or detection of FRB-like pulses from nearby GW events (discussion in Section~\ref{sec_hunt_for_bright}). 

\begin{table}
\caption{The expected sensitivity to single pulses at optimal frequency 216\,MHz (Figure~\ref{fig_sensitivity}) for different time resolutions and observing bandwidths. Assuming a typical pulse width of an FRB of 10\,ms the optimal time resolution is the same and the resulting sensitivities are 1273, 403 and 260 Jy ms for 1, 10 and 24 MWA coarse channels respectively.}
\footnotesize
\begin{tabular}{@{}cccc@{}}
\hline\hline
\textbf{Time}        & \textbf{Bandwidth} & \textbf{Sensitivity} & \textbf{Fluence}      \\
\textbf{resolution}  & \textbf{[MHz]}     & \textbf{(1$\sigma^a$)} & \textbf{10$\sigma$} \\
\textbf{[ms]}        &                    & \textbf{[Jy]}          & \textbf{threshold}  \\
                     &                    &                        & \textbf{[Jy ms]}    \\
\hline
0.1 & 1.28  & 127 & 12731 \\
    & 12.8  &  40 & 4026 \\
    & 30.72 &  26 & 2599 \\
\hline
1   & 1.28  & 40 & 4026 \\
    & 12.8  & 13 & 1273 \\
    & 30.72 & 8.2 & 822 \\
\hline
10  & 1.28  & 12.7 & 1273 \\
    & 12.8  & 4 & 403 \\
    & 30.72 & 2.6 & 260 \\
\hline
100 & 1.28  & 4.0 & 4026 \\
    & 12.8  & 1.3 & 1273 \\
    & 30.72 & 0.8 & 822 \\
\hline
\end{tabular}
\tabnote{$^a$$\sigma$ is the standard deviation of the noise}
\label{tab_sensitivity_jy}
\end{table}

\begin{table*}
\centering
\caption{List of pulsars selected for the test observations described in this paper and future work.}
\footnotesize
\begin{tabular}{@{}cccccccc@{}}
\hline\hline
\textbf{OBJECT}  & \textbf{DM}          & \textbf{W$_\text{10}$$^a$}  & \textbf{W$_\text{low}$$^b$} & \textbf{Period} & \textbf{Mean flux$^c$} & \textbf{Peak flux$^c$} & \textbf{Frequency of flux} \\
                 & \textbf{[pc/cm$^3$]} & \textbf{[ms]} & \textbf{[ms]} & \textbf{[ms]} & \textbf{density} & \textbf{density} & \textbf{density value} \\
              & &  &  & & \textbf{[Jy]} & \textbf{[Jy]} & \textbf{[MHz]}\\ %
                 
\hline%
\hline
B0950+08   & 2.97   & 20.6 & 25.3 & 253.06  & 2.37 & 29 & 150 \\ %
\hline
J0835-4510 & 67.77  & 4.5  & 50$^c$ & 89.32 & 5.9$^d$ & 10$^d$  & 215\\ 
\hline
J0837-4135 & 147.20 & 18   & 38 & 751.62  & 0.15 & 3.00 & 200 \\
\hline
J0837+0610 & 12.864 & 33.9 & 32 & 1273.77  & 1.05 & 39.5 & 150 \\
           &        &      &    &          & 0.5 & 17.5 & 200 \\
\hline
J1453-6413 & 71.248 & 9.7 & 7.1 & 179.49 & 0.63 & 11.6 & 150 \\
           &        &     &     &        & 1.2  & 22.2 & 200 \\
\hline
J1752-2806 & 50.372 & 15  & 25 & 562.56  & 1.17 & 44 & 150 \\
J1752-2806 & 50.372 & 15  & 25 & 562.56  & 2.44 & 92 & 200 \\
\hline
J0437-4715 & 2.64476 & 1.02 & 2.9 & 5.76 & 0.87 & 4.9 & 150 \\
           &         &      &     &      & 0.60 & 3.4 & 185 \\
           &         &      &     &      & 0.51 & 2.9 & 200 \\

\hline
J0630-2834 & 34.425 & 122 & 150 & 1244.42  & 0.64 & 6.5 & 150 \\
           &        &     &     &          & 0.27 & 2.8 & 200 \\

\hline
J2018+2839 & 14.1977 & 22.2 & - & 557.95  & 0.62 & 15.6 & 150 \\ 
           &         &      &     &       & 0.56 & 14.1 & 200 \\
\hline
J1456-6843 & 8.613  & 26 & 18 & 263.38  & 0.93 & 9.4 & 150 \\
           &        &    &    &         & 0.88 & 8.9 & 200 \\
\hline
B0531+21   & 56.77 & 4.7 & - & 33.39  & 7.5 & 53.3 & 150 \\ 
\hline 
\hline
\end{tabular}
\tabnote{$^a$ As in the pulsar catalogue \url{https://www.atnf.csiro.au/research/pulsar/psrcat/}}
\tabnote{$^b$ Estimates based on the earlier IC detections with the MWA by \citet{2017PASA...34...70X}, except J0835-4510.}
\tabnote{$^c$ Unless stated otherwise, mean flux density was obtained from \textsc{psrcat} at \url{https://www.atnf.csiro.au/research/pulsar/psrcat/} at the specified observing frequency.}
\tabnote{$^d$ Mean flux density based on \citep{2022PASA...39...42L}, and peak flux density
 of 10\,Jy was measured using the method described in the Appendix A and data from the Aperture Array Verification System 2 \citep[AAVS2;][]{2022JATIS...8a1014M}}
\tabnote{$^c$ Estimated using the data analysed in this paper.}
\label{tab_bright_pulsars_list}
\end{table*}



\subsection{Impact of the MWA primary beam}
We note that, although, at frequencies $\ge$200\,MHz MWA primary beam develops significant grating lobes, the sensitivity in the direction of the main lobe is not reduced by more than a factor $\sim$2 at elevations $\ge$60\degree. At these elevations optimal frequency changes only slightly (to around 180 MHz), and the sensitivity remains very close to the values in Figure~\ref{fig_sensitivity} ($\sim$1.5 -- 2\,Jy). We verified in the MWA archive that correlated observations in the frequency range 140 -- 240\,MHz at elevations $\ge$60\degree \,constitute about $80$\% of all correlated observations with the legacy MWA correlator ($\approx$73\% with new the MWAX correlator), which is a very significant fraction of observing time. Hence, we can expect that a similarly substantial fraction of observing time will be spent at these frequencies and high elevations ($\ge$60\degree), which are optimal for our FRB searches. 

Although the grating lobes at higher frequencies make the processing and potential localisations more difficult, they can provide sensitivity over larger areas of the sky. Hence, if FRBs entering the signal chain through side-lobes are sufficiently bright they can be detected and trigger recording of high-time resolution voltages. These voltages can be off-line correlated and images, including grating-lobes, can be formed as demonstrated by \citet{2021PASA...38...63C} at even higher frequencies (above 300\,MHz). Consequently such side-lobe detections could be localised, unlocking the potential of side/grating-lobe detections of very bright FRBs from low redshift Universe \citep[for example studies of side-lobe FRB detections with CHIME see][]{2023arXiv230705262L,2023arXiv230705261L}.

\subsection{Expected number of detected FRBs}
\label{subsection_expected_number_of_detected_FRBs}

\begin{figure*}[t]
\begin{center}
\vskip -0.5cm
\includegraphics[width=0.95\textwidth,angle=0]{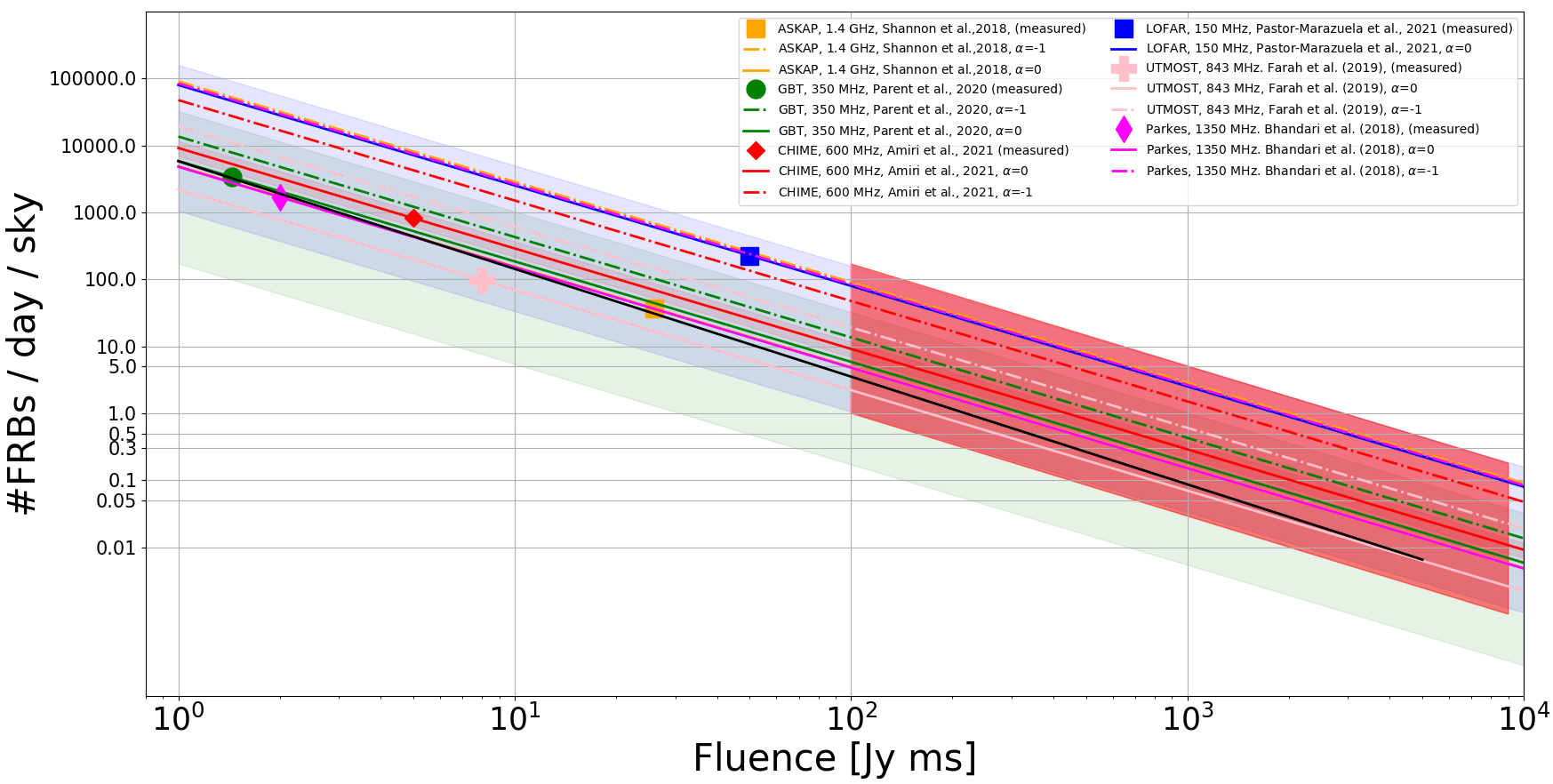}
\caption{\rednew{FRB daily rate as a function of fluence ($F$) measured by several reference instruments at frequencies from 110 to 1400\,MHz and scaled according to equation~\ref{eq_frb_rates}}. The scaling assumes Euclidean Universe (FRB rate $\propto F^{-3/2}$) which is supported by the recent CHIME results~\citep{2021ApJS..257...59C}. The measurements from ASKAP \citep{2018Natur.562..386S}, GBT \citep{2020ApJ...904...92P}, CHIME \citep{2021ApJ...923....1P}, LOFAR \citep{2021Natur.596..505P}, Parkes \citep{2018MNRAS.475.1427B} and UTMOST \citep{2019MNRAS.488.2989F} were scaled to 200\,MHz with flat spectral index, $\alpha=0$ (solid lines) and $\alpha=-1$ (dashed dotted lines), where F$\propto\nu^{\alpha}$. \rednew{The red colour marks the region with fluences $F\ge$100\,Jy\,ms 
where the FRB rate is between 0.2 and 180 per day ($\sim$360 - 65000 per year) and decreases according to $\propto F^{-3/2}$ scaling for the Euclidean Universe. This shows that the existing data from different instruments consistently predict a relatively large number ($\sim$1/day/sky) of bright low-frequency FRBs.} The much higher rate from LOFAR (blue point) was derived from the repeating FRB\,180916B during its activity period and should be treated as an upper limit. The MWA incoherent beam in 10\,ms time resolution has 10$\sigma$ detection fluence thresholds of 4000, 1300, 800 Jy\,ms for 1, 10 and 24 channels respectively (Table~\ref{tab_sensitivity_jy}). These thresholds correspond to approximately 0.02, 0.12 and 0.24 FRBs per day respectively with an uncertainty of the order of 50\% (based on the rates measured by all the different telescopes). These daily rates translate to 7, 44 and 88 FRBs per year over the entire sky for 1, 10 an 24 channels respectively and a 10$\sigma$ fluence threshold. 
However, given the FoV $\sim$ 20\degree$\times$20\degree, which corresponds to about 1\% of the entire sky, we can expect of the order of 1 FRB per year to be sufficiently bright to be detected with the described system utilising 10 or 24 MWA coarse channels. It is also clear that increasing FoV can be extremely beneficial, as an all-sky monitor described by \citet{2022aapr.confE...1S} with a detection threshold between 100 and 1000 Jy\,ms should be able to detect tens if not hundreds of FRBs per year.} 
\label{fig_frb_rates}
\end{center}
\end{figure*}

\rednew{Initially, the expected number of FRBs detected by the pipeline was estimated using the figure of merit M \citep{2004NewAR..48.1459C,2008ASPC..395..225C,2009ASPC..407..318H,Macquartetal2010} defined as:}

\begin{equation}
M = \text{FoV} \times F^{-3/2} \times \frac{\text{T}_{obs}}{\delta t},
\label{eq_fom}
\end{equation}

\rednew{where $\text{T}_{obs}$ is the total observing time, $\delta t$ is the time resolution and $F$ is the limiting fluence of a survey. This figure of merit increases with the increasing FoV, sensitivity ($F$), total observing time, and with the improved time resolution ($\delta t$). This is because instead of detecting a single pulse during observing time $T_{obs}$, the higher time resolution enables detection of multiple ($\sim T_{obs} / \delta t$) short pulses ($\le \delta t$) potentially leading to more FRB detections. In Table~\ref{tab_low_freq_searches} (9$^{th}$ column) M was normalised by $\text{M}_0$ calculated according to equation~\ref{eq_fom} for the parameters of survey by \citet{2020ApJ...904...92P} which detected one FRB at 350\,MHz.} Based on this figure of merit, the final version of the presented system with 10 coarse channels may be able to detect even $\approx$50 FRBs per year (assuming 24/7 duty cycle), which is a very optimistic prediction. However, given that the daytime data are usually unusable due to radio-frequency interference (RFI) and/or Sun power entering signal chain via side-lobes, the number of expected nighttime-only detections reduces to $\approx$25 per year. Although, observing 12 hours every day is not feasible in practise, the prediction is still quite optimistic and even allowing for 50\% downtime, the expected number will be $\gtrsim$10 FRBs per year. \red{The relatively high number of expected detections opens a possibility of significantly increasing the FRB discovery rate at frequencies $\lesssim$400\,MHz and advancing the understanding of low-frequency FRBs in general.}

\rednew{For comparison, we also estimated the FRB daily rate $R(\nu,F)$ at observing frequency $\nu$ and fluence threshold $F$ based on the reference FRB rates $R_\text{ref}$ measured by telescopes which detected many FRBs at higher frequencies (data points in Figure~\ref{fig_frb_rates}). The rate $R(\nu,F)$ was calculated according to the following equation:}

\begin{equation}
R(\nu,F) = R(\nu_{\text{ref}},F_{\text{ref}}) \times \left( \frac{\nu}{\nu_\text{ref}} \right)^\alpha \left( \frac{F}{F_\text{ref}} \right)^{-3/2},
\label{eq_frb_rates}
\end{equation}

\rednew{where $\nu_{ref}$ is the observing frequency of a reference telescope with the limiting fluence threshold $F_{ref}$, and the exponent $-\sfrac{3}{2}$ corresponds fluence scaling in the Euclidean Universe. Assuming FRB rates independent of frequency ($\alpha=0$), the expected FRB daily rate as a function of limiting fluence can be calculated at our observing frequency ($\nu$=200\,MHz) according to equation~\ref{eq_frb_rates}. Using $R(\nu_\text{ref},F_\text{ref})$ measured by other reference instruments (data points in Figure~\ref{fig_frb_rates}), the extrapolations to higher fluences are consistent in predicting} that at 10$\sigma$ fluence thresholds of 4000, 1300, 800 Jy\,ms (corresponding to 1, 10 and 24 MWA coarse channels respectively) there should be about 0.02, 0.12 and 0.24 FRBs per day per sky respectively (Table~\ref{tab_sensitivity_jy}) with an uncertainty of the order of 50\%. This corresponds to 7, 44 and 88 FRBs per year over the entire sky above the limiting 10$\sigma$ fluence thresholds for 1, 10 and 24 channels respectively. Given that the MWA FoV at 216\,MHz is $\sim$20\degree$\times$20\degree, which corresponds to about 1\% of the entire sky, we may expect of the order of 1 FRB per year to be sufficiently bright to be detected with the described system using 10 or 24 MWA coarse channels. This is about an order of magnitude less than the earlier estimate, which demonstrates the level of uncertainty of low-frequency FRB rates. Hence, one of the goals of the commensal system is to robustly establish FRB rates below 240\,MHz, which are currently poorly constrained. Furthermore, the system will be able to detect very bright FRB-like events from the local Universe (including the Milky Way galaxy), which can lead to high-impact science results.

\subsubsection{Comparison with SMART and CHASM}
\label{subsec_comparisons}

The same method can be used to estimate sensitivity of a fully coherent FRB search using 1.5\,h of MWA VCS data from the Southern-sky MWA Rapid Two-metre survey \citep[SMART;][]{2023PASA...40...21B,2023PASA...40...20B} in 140 -- 170\,MHz band, which 
reaches standard deviation of the noise $\sigma_{smart}\approx$0.5\,Jy and 1.6\,Jy in 10 and 1\,ms integrations respectively. 
This corresponds to 10$\sigma_{smart}$ fluence thresholds of 50\,Jy\,ms and 16\,Jy\,ms in 10 and 1\,ms time resolution respectively, and corresponds (based on Figure~\ref{fig_frb_rates}) to $\sim$15 and 84 FRBs per day per sky. Given that SMART observed about half of the sky for (1.5/24.0) fraction of 24-hour day, the expected numbers of FRBs detected in 10 and 1\,ms time resolution search are 0.5 and 2.6 respectively. In summary, assuming that FRB rates measured at higher frequencies can be extrapolated to MWA frequencies, of the order of 1 -- 3 FRBs can be found in SMART survey data. Such an off-line search, although computationally expensive, is currently more feasible than real-time search using \red{tied-array} beam and given its potential yield of a few FRBs is considered in the near future.

Finally, we note that as can be seen in Table~\ref{tab_low_freq_searches} the most promising low-frequency instrument to realise southern hemisphere ``FRB factory'' is an all-sky monitoring system (CHASM) described by \citet{2022aapr.confE...1S} and Sokolowski et al. (in preparation), which if implemented on SKA-Low stations can reach a detection threshold between 100 -- 1000 Jy\,ms and detect tens if not hundreds of FRBs per year.

\subsubsection{\red{Searches for transients on longer timescales}}
\label{subsec_longer_transients}

\red{The sensitivity of the incoherent beam using 24 channels (bandwidth 30.72 MHz) in 1-second time resolution is expected to be of the order of 0.3\,Jy at 210 MHz. This will be sufficient to detect longer duration dispersed radio transients. For example, like the recently discovered new class of long-period transients reported by \citet{2023Natur.619..487H,2022Natur.601..526H}, which can reach flux densities of even $\lesssim$10\,Jy. In a similar way longer-duration flares from persistent radio sources could also be detected in the incoherent sum. This will only require formation of an additional beam on a longer timescale ($\sim$0.5\,s), and sufficiently short to resolve a few second dispersion delay over $\sim$10\,MHz observing bandwidth. As in the case of FRBs, detections with presented pipeline would be verified by imaging the visibilities recorded by the standard MWA correlated mode (commensal to the presented real-time pipeline) and confirming the objects in the resulting sky images. The full discussion of this possibility is outside the scope of this paper, but given that it only requires creation of an additional lower time resolution incoherent beam it can be easily implemented and tested once the observing bandwidth is increased to $\sim$10\,MHz.}

\section{Verification of the IC pipeline}
\label{sec_verification}

\begin{figure*}[t]
\begin{center}
\vskip -0.5cm
\includegraphics[width=0.95\textwidth,angle=0]{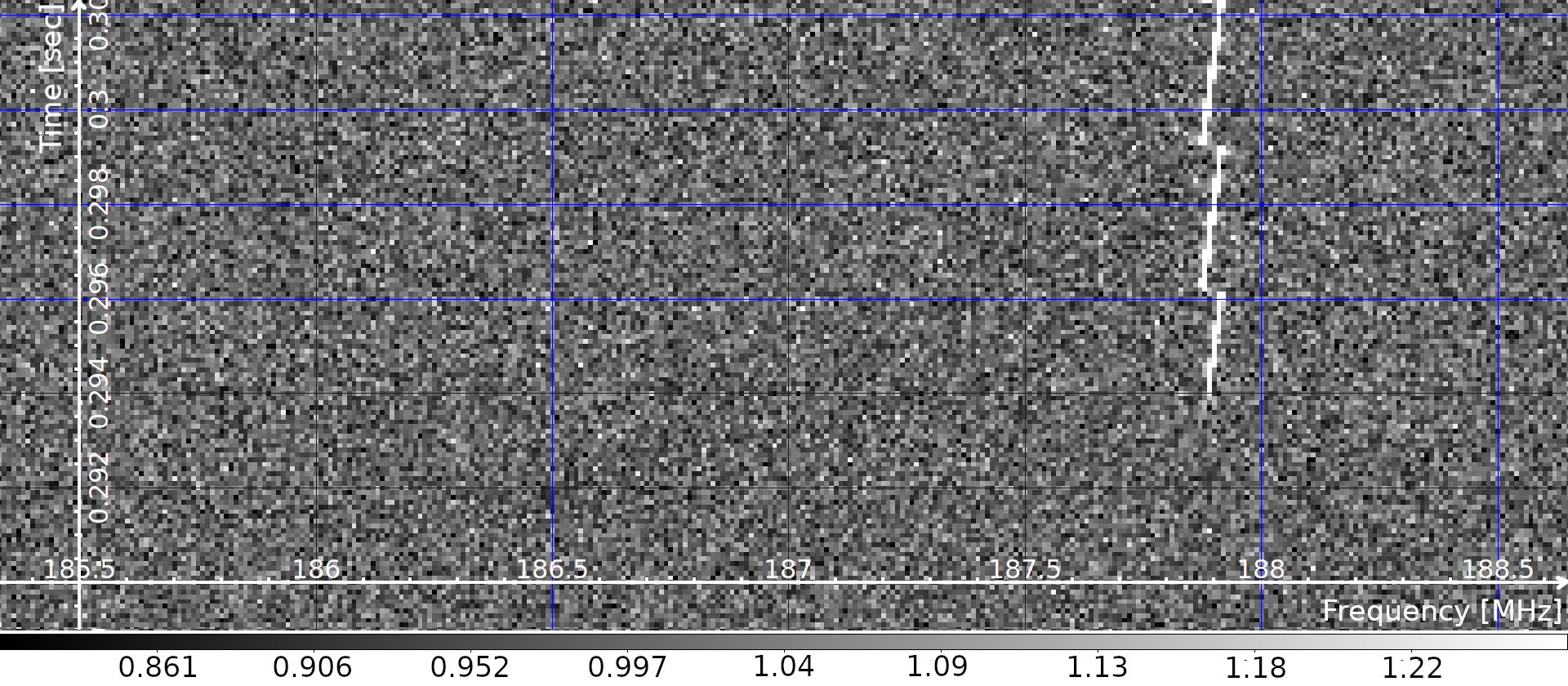}
\caption{Dynamic spectrum (frequency on X-axis and time on Y-axis) of an example radio-frequency interference (RFI) detected by the pipeline in the observation started at 2021-08-19 06:12:23 UTC in the frequency range 184.96 -- 188.80\,MHz. }
\label{fig_rfi_example}
\end{center}
\end{figure*}

The initial verification of the basic \red{mechanics} of the pipeline was achieved using \red{arbitrary} MWA observations. This confirmed that \textsc{filterbank} files were formed correctly, and could be successfully processed with FREDDA and PRESTO. These first observations were usually recorded at frequencies sub-optimal (too low) for FRB or pulsar searches, and were not targeting fields containing bright pulsars (ideal for verification of pulsar detections). Consequently, no candidates of astrophysical origin were identified, and the majority of candidates were caused by RFI (e.g. Figure~\ref{fig_rfi_example}), or abrupt changes (usually drops and less frequently \red{increases}) in total power across the entire band (1 coarse channel) due to UDP packet losses (e.g. Figure~\ref{fig_total_power}).

The best way to validate the pipeline and verify the predicted sensitivity of the search was to observe known, bright pulsars. Therefore, in the next step of the project 20\,hours of observing time per semester (so far 40\,hours in total) obtained via MWA proposal call (under the project G0086) were used to observe and verify the pipeline on selected bright pulsars. The main goals of these tests were as follows:

\begin{itemize}
\item Confirm detections of the selected pulsars by folding the resulting time series using PRESTO
\vspace{0.2cm}
\item Verify FREDDA detections of single pulses from at least some pulsars which can emit sufficiently bright single pulses (e.g. PSR B0950+08)
\vspace{0.2cm}
\item Potentially detect dispersed pulses from non-pulsar astrophysical sources
\end{itemize}
The list of the pulsars used for this verification is shown in Table~\ref{tab_bright_pulsars_list}, which includes their mean and peak flux densities, and expected and observed (if detected) SNRs of their folded pulse profiles and single pulses. A typical test recording consisted of six 300\,s or three 600\,s observations totalling to 30\,min per pulsar. 

%
%
%
\begin{table*}
\centering
\caption{List of the test pulsars used for verification of the pipeline by detection of mean profiles (of folded time series) and single pulses using an incoherent MWA beam, and one coarse channel (1.28\,MHz) at the specified observing frequency and time resolution. Mean flux densities were measured from the same MWA data or obtained from \url{https://www.atnf.csiro.au/research/pulsar/psrcat/} at the specified observing frequency. The expected SNRs were calculated using the MWA FEE beam model and pulsar parameters (Section~\ref{sec_expected_sensitivity}), and the observed SNRs for single pulses were obtained from FREDDA and for folded profiles were calculated independently of PRESTO.}
\footnotesize
\begin{tabular}{@{}cccccccccc@{}}
\hline\hline
\textbf{OBJECT}   & \textbf{Center} & \textbf{Time} & \textbf{Mean$^a$} & \textbf{Peak}  &  \textbf{Expected$^{a,b}$}  & \textbf{Expected$^{a,b}$} & \textbf{Observed} & \textbf{Observed} & \textbf{Date} \\
                  & \textbf{observed} & \textbf{Resolution} & \textbf{flux} & \textbf{flux} & \textbf{SNR} & \textbf{SNR} & \textbf{max. SNR} & \textbf{max SNR} & \textbf{of max.} \\
                 & \textbf{channels} & \textbf{[ms]} & \textbf{density} & \textbf{density} & \textbf{in avg.} & \textbf{of single} & \textbf{of avg.} & \textbf{of single} & \textbf{SNR} \\
                 & \textbf{[MHz]} &  & \textbf{[Jy]} & \textbf{[Jy]} & \textbf{profile} & \textbf{pulses} & \textbf{profile} & \textbf{pulses} & \textbf{detection} \\

\hline%
\hline
B0950+08   & 152.32        & 1  & 3.2  & 42 & 38 & 0.4 & 49 & 10 & 2023-06-01 \\ 
           & 152.32        & 1  & 3.4  & 36 & 24 & 0.3 & 37 & - & 2023-06-19 \\ 
           & 152.32        & 10 & 3.2  & 39 & 36 & 5 & 42 & 17 & 2023-01-31 \\ 
           & 152.32        & 10 & 3.2  & 39 & 36 & 5 & 42 & 17 & 2023-01-31 \\ 
           & 152.32        & 0.1 & 1.17  & 19 & 14 & 0.05 & 23 & - & 2023-06-23 \\ 

\hline
J0835-4510 & 215 & 1 & 7.8$^d$  & 31$^d$ & $\sim$10 -- 30 & $\sim$0.2 & $\approx$17 & - & 2023-06-19 \\ 
           & 215 & 10 & 4.4$^d$ &  14.4$^d$ & $\sim$10 -- 20 & $\sim$1.5 & $\approx$14 & - & 2023-02-10 \\ 
           & 215  & 100 &   7.8$^e$  & 31$^e$ & - & 1 & - & - & 2023-07-01 \\ 
\hline
J1752-2806$^c$ & 215 & 1 & 2.4$^f$    & 92   & 38.0 & 1.20 & 7 & - & 2023-06-01 \\ 
               & 215 & 1 & 0.8    & 25   & 6 & 0.1 & 8.5 & - & 2023-06-19 \\ 
               & 152.32 & 1  & 1.17$^f$   & 44   & 10.4 & 0.34 & 5 & - & 2023-06-01 \\ 
\hline
 J1456-6843 & 150 & 1 & 1.06 & 18 & 4 & 0.03 & 12 & - & 2023-07-12 \\ %
\hline
J0837-4135 & 215  &  1 & 0.15$^f$ & 6.30 & 7.7  & 0.1  & - & - & 2023-06-22 \\ 
           & 121  &  1 & 0.15$^f$ & 6.30 & 4.9  & 0.05 & - & - & 2023-06-22 \\ 
\hline
J0837+0610 & 215 &  1  & 0.4  & 8 & 10 & 0.2 & 6 & - & 2023-06-20 \\ 

\hline
J1453–6413 & 152.32 & 0.1 & 0.63$^f$    & 11.6 & 3.6 & 0.1 & 3 & - & 2023-06-23 \\ 
           & 215    & 0.1 & 1.2$^f$     & 23 & 9.6 & 0.1 & 3.5 & - & 2023-07-12 \\ 
\hline
J0437-4715 & 152.32 & 0.1 & 0.87$^f$ & 4.9 & 7 & 0.02 & - & - & 2023-06-24 \\ %
               & 215 & 0.1 & 0.50$^f$ & 2.9 & 4.4 & 0.02 & - & - & 2023-06-24 \\ %
\hline
B0531+21   &  152.32           & 0.1    & 7.5$^f$ & 53.3  & 24 & 0.08 & - & - & 2023-06-24 \\ 
           &  215              & 0.1    & 7.5$^f$ & 53.3  & 27 & 0.09 & - & - & 2023-06-24 \\ 

\hline 
\hline
\end{tabular}
\begin{flushleft}
\tabnote{$^a$ If the pulsar was detected and not indicated explicitly with $^f$, we provide the mean flux density measured ($f_{meas}$) with the method described in the Appendix A applied to this particular data, and otherwise we provide the value ($f_{cat}$) from \textsc{psrcat}. Consequently, the expected SNR values were calculated using mean flux densities in this column. Hence, in order to calculate expected SNRs for the mean flux densities in \textsc{psrcat} a scaling factor $f_{cat} / f_{meas}$ has to be applied. For example, for pulsar B0950+08 catalogue flux density of 2.37\,Jy, peak flux is 29\,Jy, while expected SNRs of folded profile is 26.}
\tabnote{$^b$ The expected SNRs are approximate and dependent on pulse width and mean flux density, which are not always well known at these frequencies (or can vary). If a pulsar was detected its flux density was usually estimated from its mean pulse profile.}
\tabnote{$^c$ Peak flux density and SNRs are lower by a factor $\approx$1.67 if the width is estimated from \citet{2017PASA...34...70X} measurements at 185\,MHz.}
\tabnote{$^d$ Measured (see Appendix A for the method) on a particular date using the exact same data and they resulted in the upper end of SNR range (SNR 30 and 20 for 1 and 10ms time resolutions). }
\tabnote{$^e$ Used the same values as measured from 2023-06-19 data as we cannot detect Vela with 100\,ms time resolution given its period of $\approx$89.3\,ms.}
\tabnote{$^f$ Flux density value from the ATNF pulsar catalogue, and otherwise it was measured from the data}
\end{flushleft}
\label{tab_bright_pulsars}
\end{table*}

\red{As mentioned in Sections~\ref{subsec_frb_searches_below_350MHz} and \ref{subsection_expected_number_of_detected_FRBs}, nighttime observations were preferred. There are two main reasons for this. Firstly, there is much more RFI during daytime due to ongoing maintenance and other work at and around the Observatory. Secondly, due to the impact of the Sun. Both RFI and the Sun can enter the signal path via the main beam or side-lobes. However, typical daytime MWA observations are pointed away from the Sun, so it is more likely that radio waves from the Sun (often variable) enter the signal path via side-lobes, which is worse as it is practically impossible to remove (deconvolve) from images (due to uncertainties in the beam model and flux density of the Sun at the time of observations). In the case of incoherent beam RFI and Sun signals (both can be highly variable) entering via the main lobe or side-lobes of the primary (tile) beam can cause spurious effects, artefacts and false positive candidates. Therefore, daytime data quality are usually much lower and such observations are generally avoided when high quality data are required (they can still be used for initial testing).}

\subsection{Preliminary results}
\label{sec_results}

As described in Section~\ref{sec_verification} the pipeline has been verified by observing selected bright pulsars listed in Table~\ref{tab_bright_pulsars}, where columns 5 and 6 show the expected SNRs of the averaged profile and single pulses respectively, while columns 7 and 8 show the observed SNRs if a particular pulsar was detected. The selection of these pulsars was based on the early MWA census using an offline incoherent beam pipeline \citep{2017PASA...34...70X} and observations with SKA-Low stations \citep{2021PASA...38...23S,2022PASA...39...42L}. This section provides a discussion of pulsar detections and explanations of the non-detections. 

\subsubsection{B0950+08}

\begin{figure*}[t]
\begin{center}
\includegraphics[width=0.95\textwidth,angle=0]{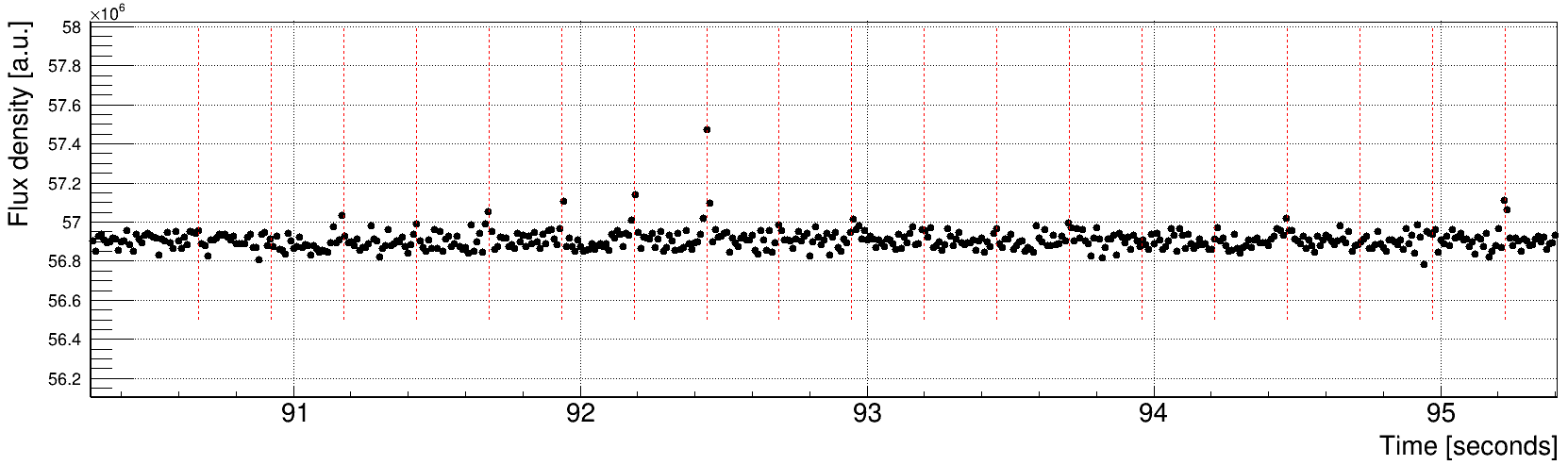}
\caption{Time series from the 2023-02-01/02 observation of PSR B0950+08 dedispersed with DM=2.9\,pc/cm$^3$. The black points are the observed data and red dashed lines separated by the pulsar period ($\approx$0.253\,s) mark the expected pulse arrival times. A very bright pulse with SNR$\sim$15 at $\approx$92.44\,s since the start of the observation is clearly visible together with several fainter pulses. The corresponding dynamic spectrum is shown in Figure~\ref{fig_b0950_single_pulses}.}
\label{fig_b0950_single_pulses_timeseries}
\end{center}
\end{figure*}
\begin{figure*}[t]
\begin{center}
\includegraphics[width=0.95\textwidth,angle=0]{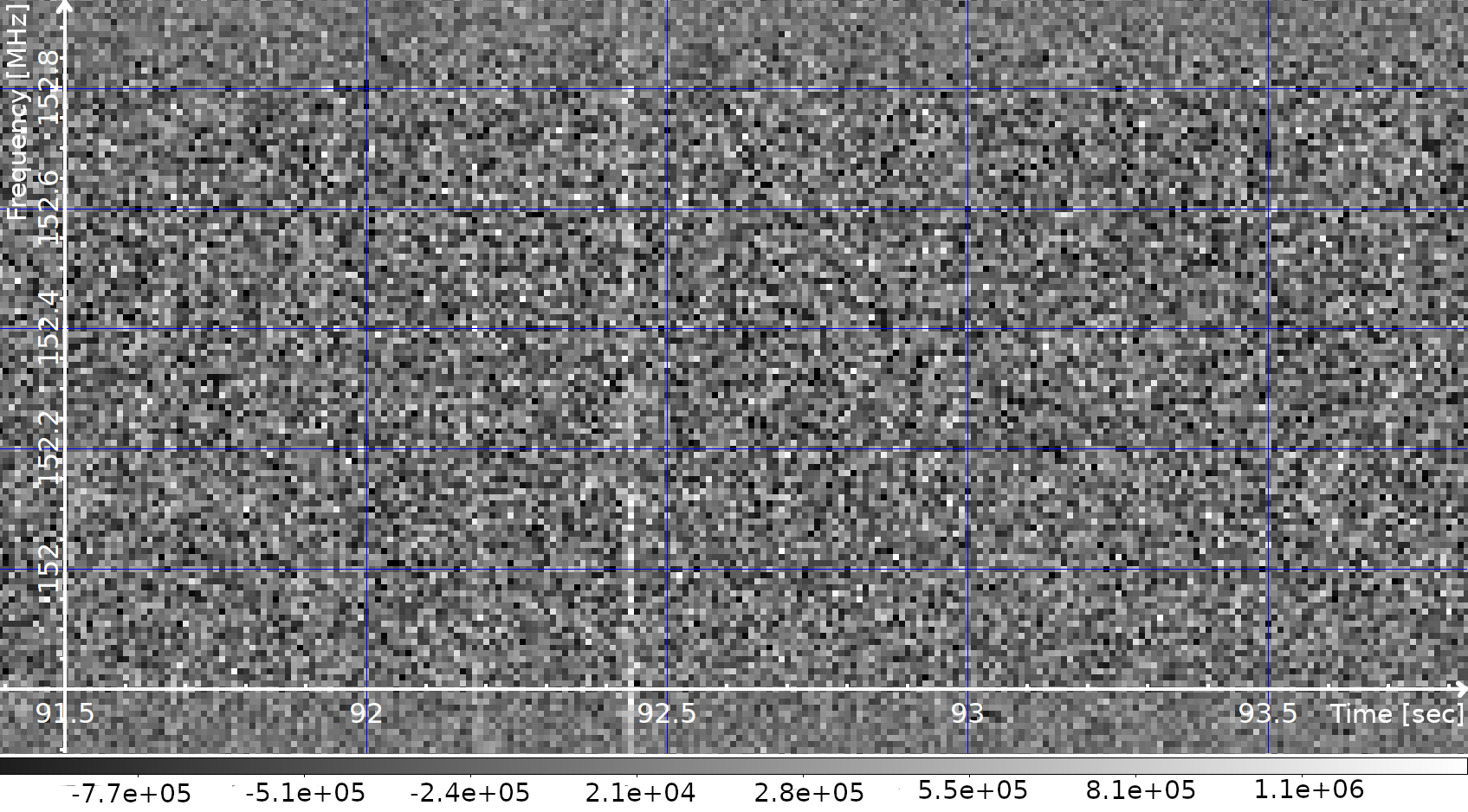}
\caption{Dynamic spectrum after subtraction of the mean bandpass. The X axis is time in 10\,ms resolution, the Y axis is frequency in 10\,kHz resolution, and the colour scale is flux density (in arbitrary units). Two pulses from pulsar B0950+08 are clearly visible. The brighter at approximately 92.44 seconds since the start of the observation, and a fainter pulse one pulsar period ($\sim$0.253\,s) earlier. Upon careful inspection it can be seen that the bright pulse arrives by about 1 pixel (timestep of 10ms) earlier at higher frequency (154.24\,MHz at the top of the image) and later at lower frequency (152.96\,MHz at the bottom of the image), which agrees with the expected dispersion delay of about 8\,ms for this frequency range and pulsar DM = 2.97 pc/cm$^3$. The dynamic spectrum shows the full coarse channel. The corresponding de-dispersed time series is shown in Figure~\ref{fig_b0950_single_pulses_timeseries}.}
\label{fig_b0950_single_pulses}
\end{center}
\end{figure*}
\begin{figure*} 
\begin{center}
\includegraphics[width=0.95\textwidth,angle=0]{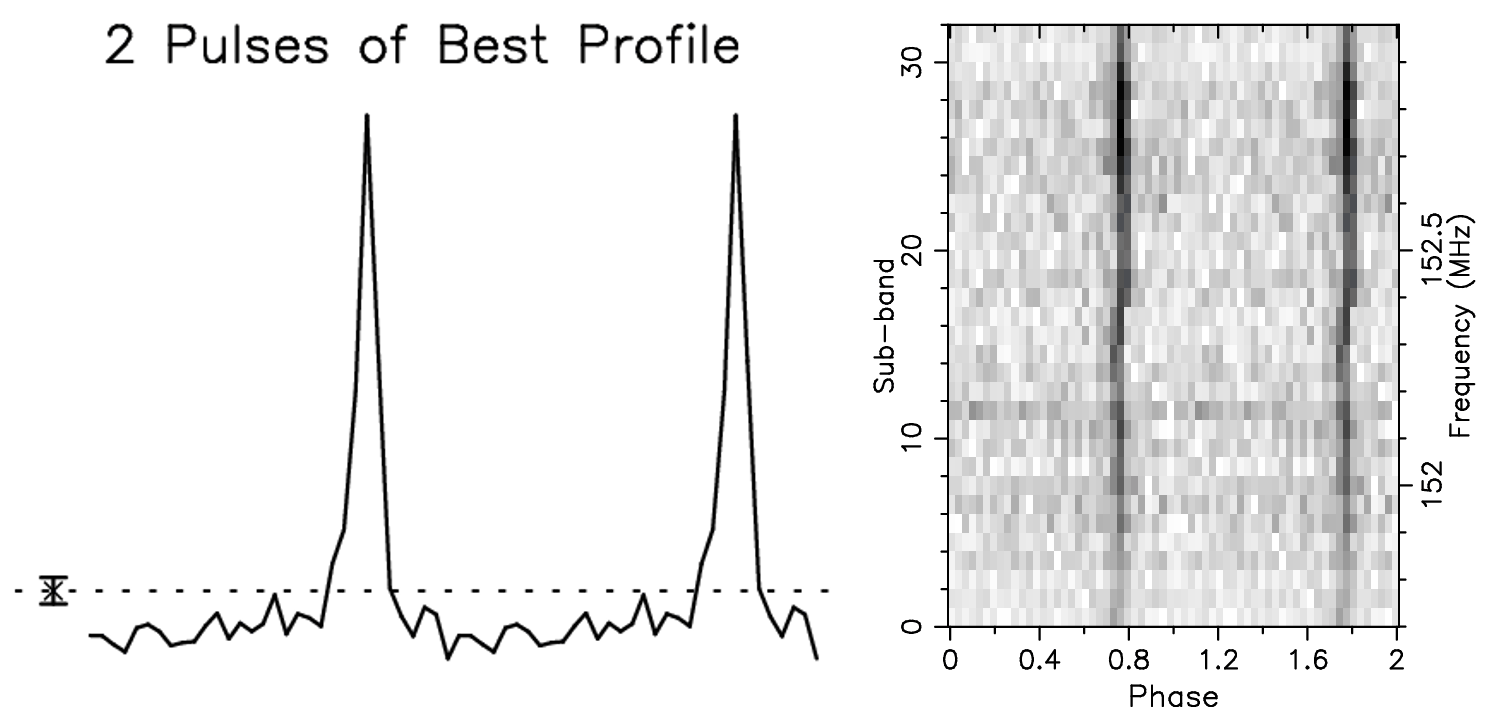}
\caption{Folded profile of pulsar B0950+08 obtained from 5\,min of data. The observation was started on 2023-06-01 10:14:47 
UTC recording with 1\,ms time resolution. \textit{Left}: The mean profile of the pulsar. \textit{Right}: The dynamic spectrum (frequency vs. phase). The PRESTO SNR of this detection was 44.6, while the SNR estimated independently of PRESTO was $\approx$49.}
\label{fig_b0950_folded}
\end{center}
\end{figure*}

Pulsar PSR B0950+08 is a well-known pulsar first reported as Cambridge Pulsed source CP0950 by \citep{1968Natur.218..126P}. Its proximity (DM $\approx$ 2.97 pc/cm$^3$) makes it highly variable due to refractive and diffractive scintillation events, and individual pulses can reach flux densities as high as $\approx$155\,Jy as observed in all-sky images from the SKA-Low prototype stations \citep{2021PASA...38...23S}, MWA images \citep{2016MNRAS.461..908B} and by other instruments~\citep{2020MNRAS.497..846K}. It is a bright pulsar with mean flux density $\approx$2.4\,Jy at 150\,MHz (Table~\ref{tab_bright_pulsars_list}), which makes it is one of the best test pulsars for the verification of the presented pipeline both by detection of mean pulse profiles and single pulses. 
The average pulse profile of B0950+08 was detected in all observations in the frequency band 140 -- 170 MHz with SNR up to 49 in 1\,ms time resolution. It was also the only object from which single pulses were detected with SNR $\approx$ 15.5, but this was evident only in 2 datasets when multiple single pulses were detected with time separations consistent with the pulsar period P$\approx$253\,ms. It is worth noting that it is possible that single pulses were detected in more observations, but due to the very low DM of this pulsar they are often indistinguishable from RFI unless multiple pulses are detected and readily separated by the pulsar period. Example single pulse detections are shown in Figures~\ref{fig_b0950_single_pulses_timeseries}, and \ref{fig_b0950_single_pulses}, and 
mean pulse profile in Figure~\ref{fig_b0950_folded}.

\subsubsection{J0835-4510 (Vela)}

\begin{figure*} 
\begin{center}
\vskip -0.5cm
\includegraphics[width=0.95\textwidth,angle=0]{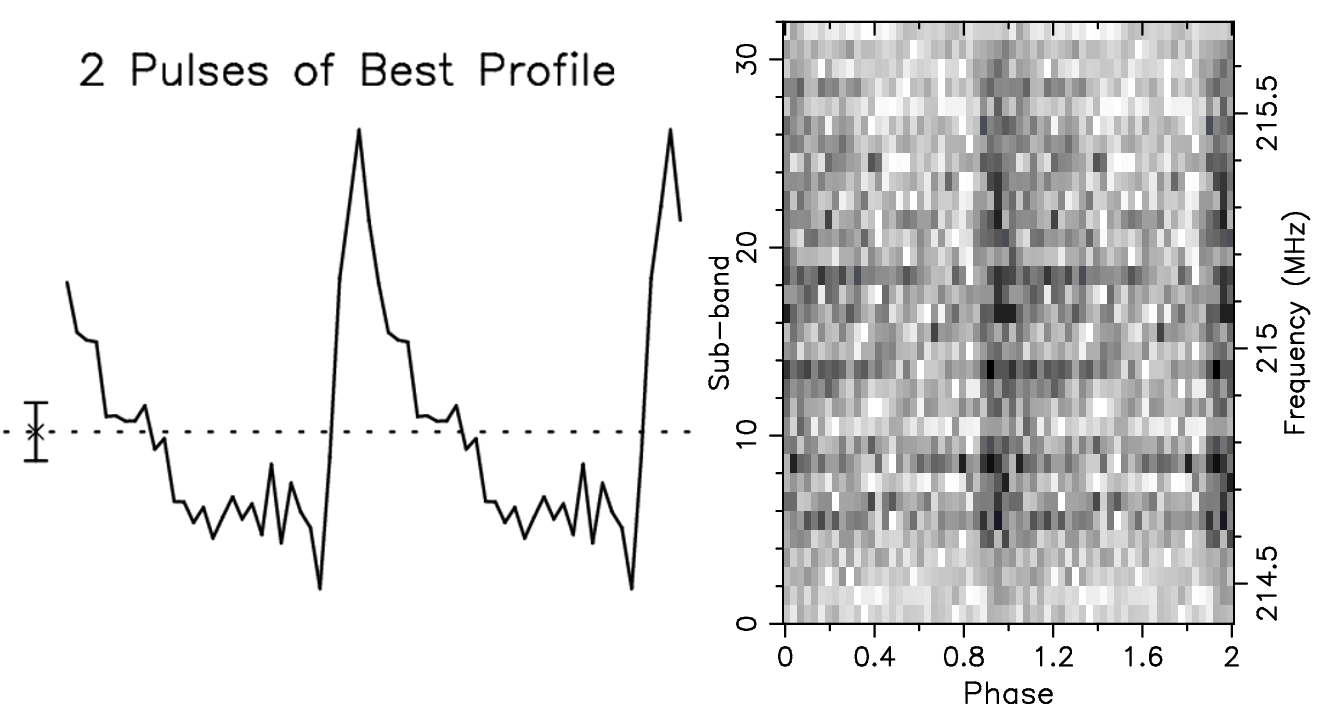}
\caption{Folded profile of pulsar J0835-4510 (Vela) obtained from 5\,min of data. The observation was started on 2023-06-19 06:49:59 UTC recording with 1\,ms time resolution. \textit{Left}: The mean profile of the pulsar. \textit{Right}: The dynamic spectrum (frequency vs. phase). PRESTO SNR of this detection was $\approx$23, while SNR estimated independently of PRESTO was $\approx$17.}
\label{fig_vela_folded}
\end{center}
\end{figure*}

PSR~J0835-4510, also known as the Vela pulsar, is one of the best known and studied pulsars, discovered by \citet{1968Natur.220..340L}. It was the first ever association of a supernova remnant and a pulsar, and has a very high mean flux density of $\approx$6\,Jy at 215\,MHz \citep{2022PASA...39...42L}, which makes it an ideal probe for verification of the pipeline, telescope performance and data quality. Therefore, it was decided early on during the observing campaign to always record at least $30$\,min of Vela data. The detection of its mean profile was used as a verification of the performance of the entire system. In order to minimise the effects of scattering, these test observations were performed mostly in 200 -- 230\,MHz frequency range as a compromise between scattering and sensitivity of the MWA (which is degraded at frequencies above 280\,MHz). As expected (Table~\ref{tab_bright_pulsars}) Vela's average profile was always detected with SNR$\sim$15 (example in Figure~\ref{fig_vela_folded}). However, also in-line with the expectations, no single pulses from Vela were detected. Since the flux densities of pulsars can change on daily timescales, and Vela is affected by multipath scattering due to its location within the supernova remnant, an additional cross-check of the detected flux densities was performed with the same data using the method described in Appendix A. This method is robust against scattering, but otherwise is very similar to the one used by \citet{2022PASA...39...42L}. It was also applied to other detections of mean profiles in order to estimate the pulsar flux densities on a given day and for a specific observation. 

\medskip 
%

\begin{figure*} 
\begin{center}
\includegraphics[width=0.95\textwidth,angle=0]{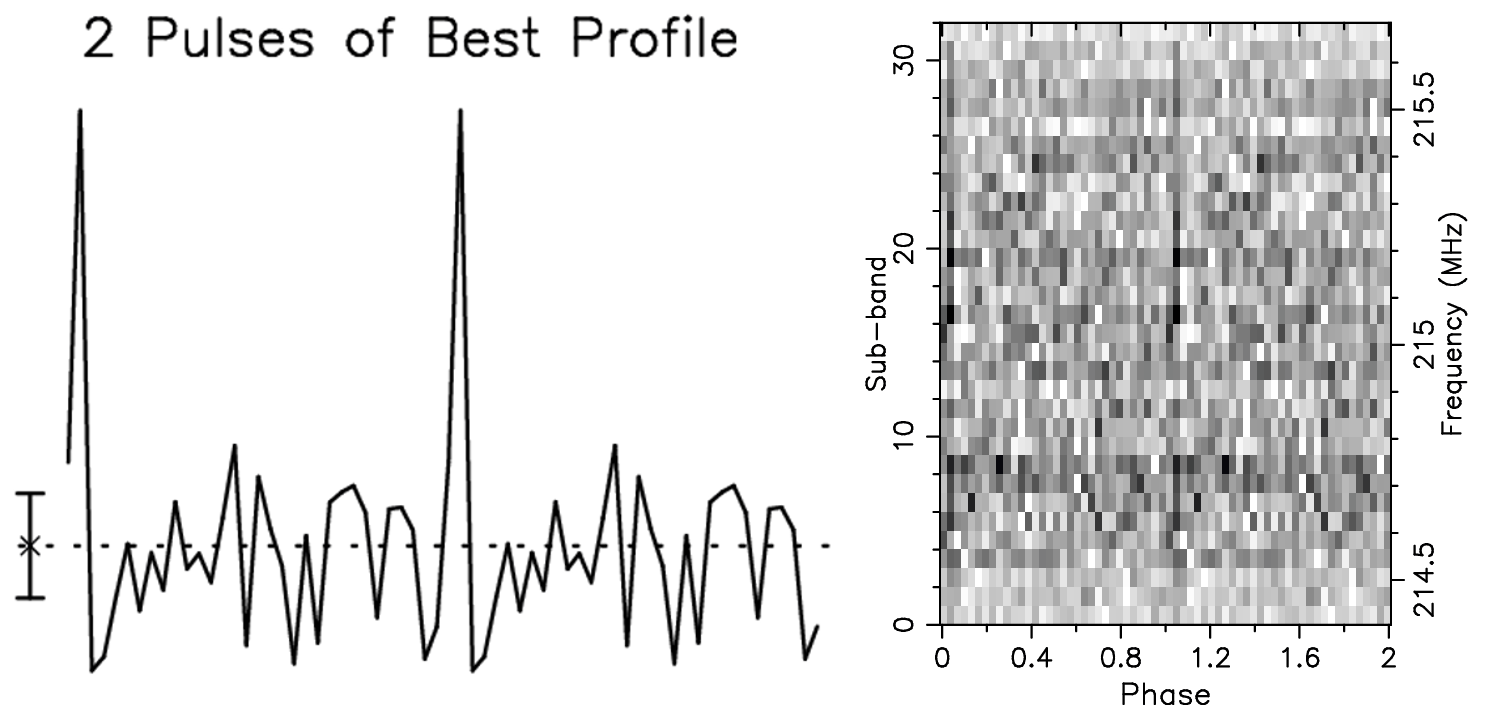}
\caption{Folded profile of pulsar J1752-2806 obtained from 190\,sec of data. The observation was started on 2023-06-19 16:04:55 UTC recording with 1\,ms time resolution. \textit{Left}: The mean profile of the pulsar. \textit{Right}: The dynamic spectrum (frequency vs. phase). The SNR of this detection was $\approx$7 (both via PRESTO and independently of PRESTO).}
\label{fig_j1752_folded}
\end{center}
\end{figure*}

\begin{figure*} 
\begin{center}
\includegraphics[width=0.95\textwidth,angle=0]{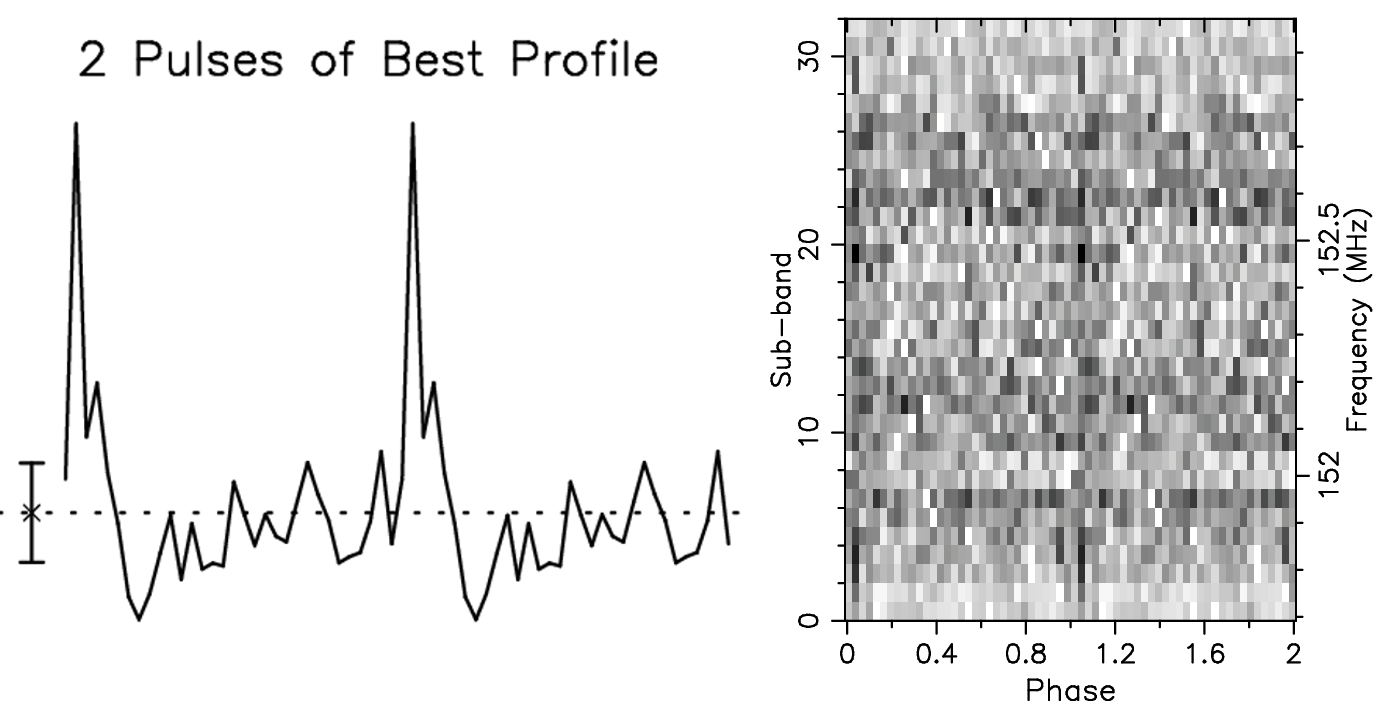}
\caption{Folded profile of pulsar J1456-6843 obtained from 480\,sec of data. The observation was started on 2023-07-12 12:35:03 UTC in recording with 1\,ms time resolution. The mean profile of the pulsar. \textit{Right}: The dynamic spectrum (frequency vs. phase). The PRESTO SNR of this detection was 5.7, while the SNR estimated independently of PRESTO was $\approx$12.}
\label{fig_j1456m68_folded}
\end{center}
\end{figure*}

\subsubsection{J1752-2806}
Pulsar PSR J1752-2806 is another very bright pulsar with mean flux density $\sim$2\,Jy. Its folded profile was detected with maximum SNR$\approx$8 (Figure~\ref{fig_j1752_folded}). It was a factor of a few ($\sim$2 -- 3) lower than the expected SNR, and the reasons for such a lower observed SNR are still investigated. As expected (Table ~\ref{tab_bright_pulsars}) no single pulses from this pulsar were detected.

\subsubsection{Detections of other pulsars}

Besides the three pulsars listed above several other pulsars listed in Tables~\ref{tab_bright_pulsars_list}~and~\ref{tab_bright_pulsars} were also observed, and the following pulsars were marginally detected with very low SNRs:

\begin{itemize}
   \item J1456-6843 detected at 150\,MHz in 1\,ms time resolution (Figure~\ref{fig_j1456m68_folded}) and PRESTO SNR$\approx$6 (SNR estimated independently of PRESTO $\approx$ 12).

   \item J1453-6413 marginal detection at 215\,MHz with SNR $\approx$3.5 (2023-07-12), and at 152.32 MHz with SNR $\approx$3 (2023-06-23). Both in 1\,ms time resolution. 
\end{itemize}

No single pulses from pulsars other than B0950+08 were detected with FREDDA, which mostly agrees with our expectations for the sensitivity of the current search using a single coarse channel (see Table~\ref{tab_bright_pulsars}). Unfortunately, several pulsars (J0837+0610, B0531+21, J0437-4715 and J0837-4135) could only be observed during daytime, which is most likely the main reason for their non detections. Besides daytime observations, some of these non-detections could be caused by non-favourable \red{interstellar weather} conditions on the days of observations as even B0950+08 was not always bright enough to be unanimously detected in single pulse searches. Another reason for the non-detections may be the increased system noise due to contributions from malfunctioning (or broken) tiles as flagging and exclusion of such tiles is not implemented in the currently tested version of the real-time beamformer software (it is planned in the next versions of the code). The system presented here used only a single MWA coarse channel (1.28\,MHz), which resulted in very limited sensitivity. Thus, non-detections of some pulsars are not unexpected. Especially, that many of them could only be observed during daytime. However, as the system is upgraded with larger observing bandwidth, the testing will continue, and the non-detected pulsars will be re-observed during nighttime.



\subsection{Candidates due to instrumental effects and RFI}

Besides detections of astrophysical objects the pipeline identified candidates caused by ``closer to Earth'' effects, which will be briefly summarised here. 

\subsubsection{Radio-frequency interference (RFI)}

The majority of initial tests were performed using existing MWA observations (typically 120\,s) scheduled for various approved projects and performed at \red{arbitrary} frequencies. During these observations all the candidates identified by the pipeline were attributed to various forms of RFI (example in Figure~\ref{fig_rfi_example}). 

\subsubsection{Variations in power of the IC beam}

\begin{figure*}[t]
\begin{center}
\includegraphics[width=0.95\textwidth,angle=0]{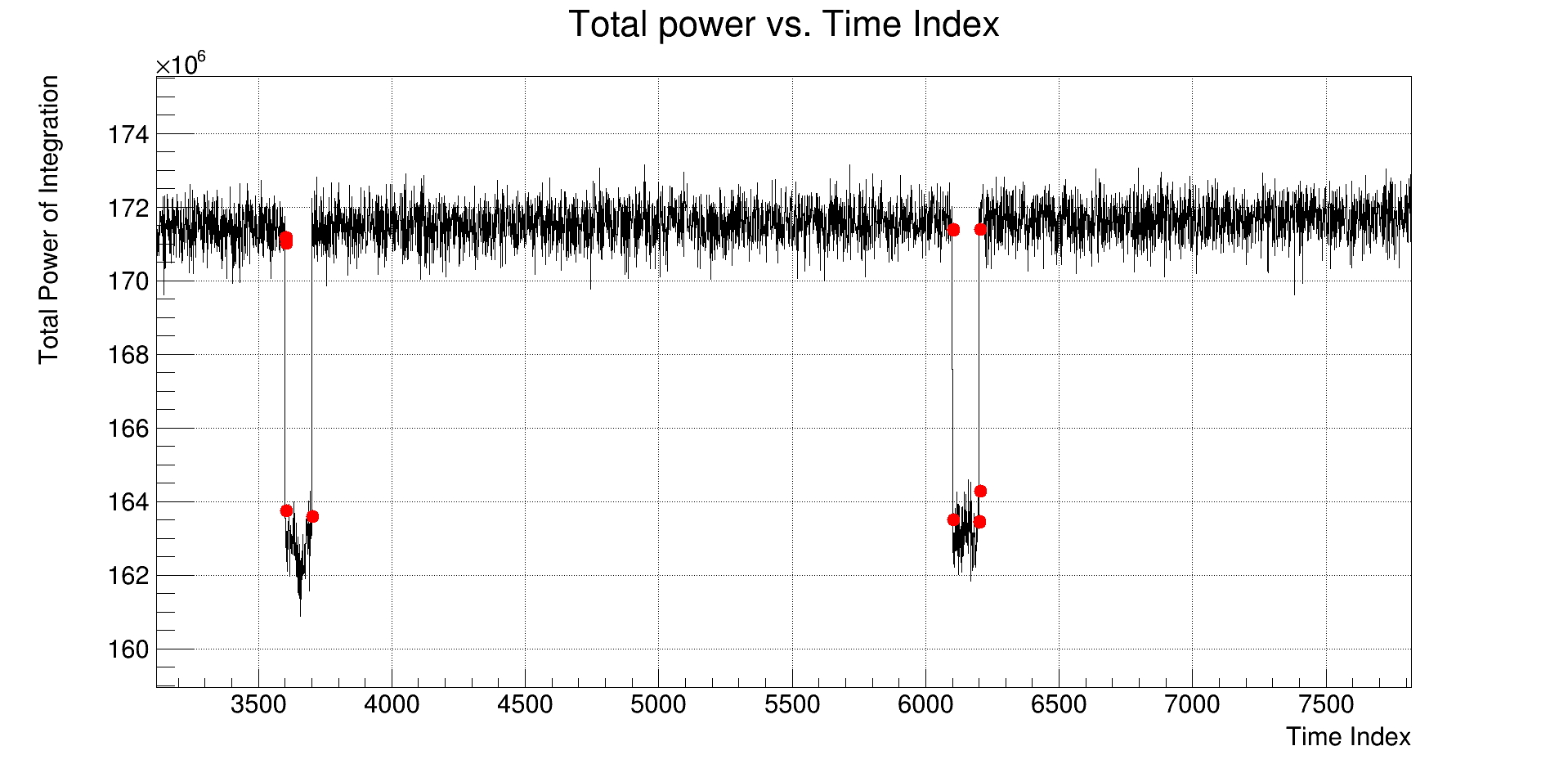}
\caption{Total power as a function of time sample index from one of the Vela observations (internal MWA ID 1359388216). The black curve is the total power in a single coarse channel (153.6\,MHz in this case), and the red dots indicate the drops in total power as automatically detected by our software. A single time step corresponds to 10\,ms. Hence, the first drop in power (i.e. loss of packets) occurred during a 1-second block started 36\,sec after the start of the observation.}
\label{fig_total_power}
\end{center}
\end{figure*}

One of the most common sources of false-positive candidates from FREDDA were abrupt changes of power which can be seen in the dynamic spectrum and total power calculated over a single coarse channel (example in Figure~\ref{fig_total_power}). These power jumps were observed during nearly all observations, and therefore a simple \red{code detecting abrupt changes in total power} was developed to identify this kind of events and excise 
 candidates caused by them. An example of total power as a function of time during one of the 30\,min observations with automatically detected power jumps marked with red dots is shown in Figure~\ref{fig_total_power}. These total power jumps were attributed to UDP packet losses, which caused certain portions of data being lost, that is not included into the IC sum and consequently reducing the observed total power of the incoherent beam. The times when they occurred turned out to be exactly matching the times of packet loses recorded in MWA log files, which confirmed the initial hypothesis. This observation triggered software improvement work, which will hopefully minimise these effects and its negative impact on IC pipeline in the future. 

 On the other hand, similar power variations were observed in V-FASTR experiment \citep{2011ApJ...735...97W} and were resolved on the data processing level by subtracting a running mean of signal power. This removed power jumps in their data, and may also be the best way to improve the real-time MWA IC pipeline, as even after further software/hardware improvements there may always be some small residual packet losses, and the FRB/SETI search software should be robust against these situations.

\section{Summary and future plans}
\label{sec_summary_future}

The commensal pipeline for FRB and SETI searches is currently in early stages of development, and can form several beams using a single MWA coarse channel (1.28\,MHz). The mechanics of the pipeline was verified, and confirmed to produce \textsc{filterbank} files commensally to on-going MWA observations. We have also deployed a first version of FRB search pipeline running FREDDA software on these filterbank files. The commensal pipeline was also verified on selected pulsars using dedicated observing time (total 40\,h). Mean profiles of pulsars B0950+08 and Vela were detected with maximum SNRs $\approx$49 and $\sim$17 respectively. However, single pulses were unequivocally identified only in several observations of the pulsar PSR B0950+08 (maximum SNR $\approx$ 17). These results mostly agree with the expectations (Table~\ref{tab_bright_pulsars}). Especially, that some of the pulsars (e.g. J0837+0610 or PSR B0531+21), which in principle should (at least occasionally) have sufficiently bright single pulses to be detected by the pipeline, could only be observed during daytime making their detections highly unlikely. These objects will be re-observed during nighttime and possibly with larger observing bandwidth. Another possible reason for the non-detections may be a few bad tiles, which are not excluded from the incoherent beam (i.e. sum), contribute noise and possibly RFI, and make detections of fainter pulsars impossible. The exclusion of flagged tiles will be included in the next version of the code.

The results of the presented tests show that FREDDA software can be successfully applied to the incoherent MWA beam as it was able to identify most of the pulses from PSR B0950+08. Using one of the datasets with multiple bright pulses, it was verified that $\approx$85\% of them were identified by FREDDA. So, far we were not able to detect single pulses from higher DM objects, but this will become possible once the observing bandwidth is increased. The system is currently being upgraded with additional computers, which should enable capturing and processing of a few coarse channels (up to 10 corresponding to 12.8\,MHz bandwidth) leading to an increase in the sensitivity of the FRB search by a factor of $\approx$3. The main source of false-positive candidates identified by FREDDA were \red{the abrupt changes in total power}, which will be improved by reducing the number of lost packets (ongoing software work) and/or by more robust data processing (e.g. subtraction of running mean). Moreover, addition of an automatic candidate classification is also planned in the near future. 

Verification of commensal coherent beamforming functionality is another short term goal, which will enable recording of \red{tied-array} beams on several objects within the MWA primary beam. It will speed-up processing of MWA high-time resolution data enabling routine pulsar timing observations, and monitoring of known repeating FRBs. Furthermore, it will also significantly reduce the storage requirements as \red{tied-array} beam for a single object requires $N_{ant}$ (128 or 256) less disk space than the full MWA VCS data product. The presented commensal pipeline is routinely running during most of the MWA observations, and once the frequency bandwidth is increased may be able to detect a several bright FRBs per year. 


\begin{acknowledgements}
This scientific work uses data obtained from Inyarrimanha Ilgari Bundara / the Murchison Radio-astronomy Observatory. We acknowledge the Wajarri Yamaji People as the Traditional Owners and native title holders of the Observatory site. Support for the operation of the MWA is provided by the Australian Government (NCRIS), under a contract to Curtin University administered by Astronomy Australia Limited. We acknowledge the Pawsey Supercomputing Centre which is supported by the Western Australian and Australian Governments. Inyarrimanha Ilgari Bundara, the CSIRO Murchison Radio-astronomy Observatory and the Pawsey Supercomputing Research Centre are initiatives of the Australian Government, with support from the Government of Western Australia and the Science and Industry Endowment Fund. This research has made use of NASA's Astrophysics Data System. 
\end{acknowledgements}

\vspace{0.5cm}

\textbf{Appendix A}
\begin{appendix}
\label{appendix_a}

\begin{figure}[t]
\begin{center}
\includegraphics[width=0.45\textwidth,angle=0]{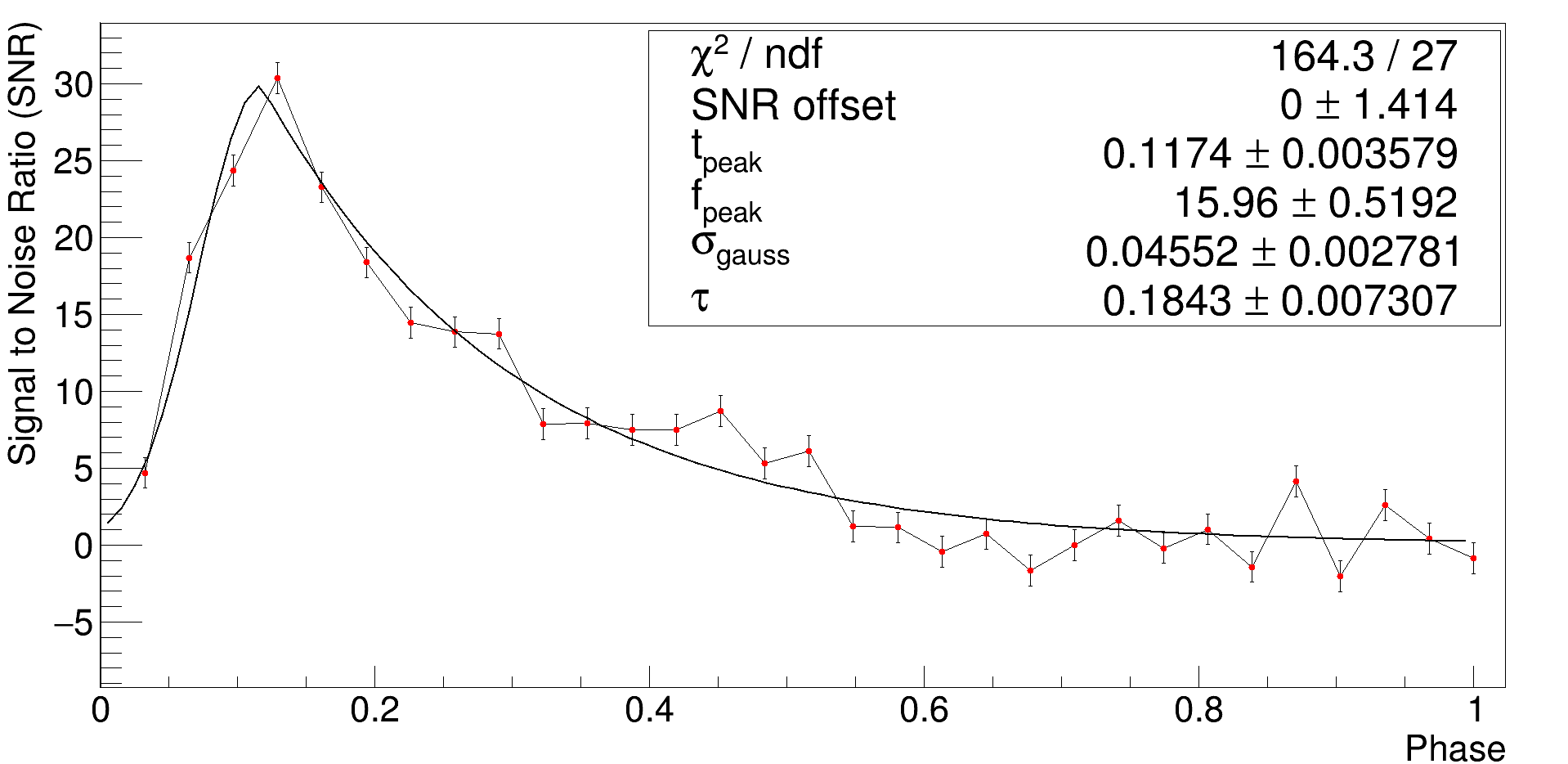}
\caption{Example fit of equation~\ref{eq_vela_pulse_func} to 2013-06-14 MWA data at 216\,MHz, which enabled calculation of the mean flux m$\approx$7.8\,Jy and peak flux$\approx$31\,Jy, and estimation of W10$\approx$50\,ms as the time width of the pulse where the flux density is $\ge$10\% of the peak value (summarised in Table~\ref{tab_bright_pulsars}) .}
\label{fit_vela_pulse_fit}
\end{center}
\end{figure}

The flux density of a pulsar can be estimated from folded profile created by pulsar processing software such as \textsc{PRESTO} or \textsc{DSPSR}, and System Equivalent Flux Density (SEFD) calculated using beam model of an MWA tile or an SKA-Low station \citep{2022PASA...39...15S}. This approach was successfully applied to SKA-Low prototype stations data to measure flux densities of selected pulsars as described by \citet{2022PASA...39...42L}. In this paper we used the same method to measure mean flux density of Vela pulsar in data from the Aperture Verification System 2 (AAVS2), and we also applied similar procedure using MWA Full Embedded Element (FEE) beam model \citep{2017PASA...34...62S} to the MWA data recorded for the work presented here. Vela profile is highly scattered at low frequencies (example in Figure~\ref{fit_vela_pulse_fit}) and it is often difficult to find a suitable window within the pulse period when the pulsar emission is not present and observed standard deviation ($\sigma_o$) of the noise can be calculated. However, either the phase bins before the on-set of the pulse can be used, or a mathematical representation of the pulse profile can be fitted, subtracted and $\sigma_o$ can be measured as the standard deviation of the residuals. We used both these approaches and they led to nearly identical results. 
In a short summary mean flux and peak flux densities of Vela pulsar and other pulsars in our test sample were calculated according to the following procedure:
\begin{enumerate}
\item Calculate observed standard deviation of the noise ($\sigma_o$) from the off-pulse part of the folded profile (e.g. phase window 0.8 - 1.0 in Figure~\ref{fit_vela_pulse_fit}) or from the residuals after subtracting the fitted mathematical representation of the mean profile.

\item Divide average pulse profile by $\sigma_o$ in order to re-scale uncalibrated flux density such that new standard deviation is 1. Divide values on X-axis by their maximum value (typically number of bins $n_{bin}$ if it was originally bin index) in order to re-scale X-axis to [0,1] range. 

\item Fit a mathematical function to the mean pulse profile. We found that the best fit resulted from a sum of a Gaussian function (representing the on-set and peak of the pulse) multiplied by exponential decay (only for phase values higher than phase of the peak flux). This can be mathematically described as:

\begin{equation}
f(p) = 
\begin{cases}
 G(p)\text{ for p }\le p_{peak} \\
 G(p) \cdot e^{-\frac{(p-p_{peak})}{\tau}} \text{ for p }\ge p_{peak} \\
\end{cases}
\label{eq_vela_pulse_func}
\end{equation}

where $G(p)$ is the Gaussian profile used to describe the pulse peak:

\begin{equation}
G(p) = \frac{f_p}{\sqrt{2\pi \sigma_p}} \cdot e^{-\frac{(p-p_{peak})^2}{2\sigma_p^2}},
\end{equation}

and p is the phase of the average pulse period in [0,1] range, $\sigma_p$ is standard deviation of the Gaussian profile, $p_{peak}$ is the phase corresponding to peak flux density, $\tau$ corresponds to scattering time in units of pulsar phase (hence unitless here) and $f_p$ is the peak flux density of the average profile. Example of fitted profile is shown in Figure~\ref{fit_vela_pulse_fit}. Simpler pulse profiles (e.g. for pulsar B0950+08) were fitted with a Gaussian profile ($G(p)$) only, i.e. without the exponential \red{tail} term.

\item Calculate expected standard deviation of the noise ($\sigma_s$) in the incoherent sum of all MWA tiles, using $\text{SEFD}_X$ and $\text{SEFD}_Y$ calculated with the MWA FEE Beam model \citep{2017PASA...34...62S} and the following equation:

\begin{equation}
\sigma_s = \frac{\text{SEFD}_I\sqrt{n_{bin}}}{\sqrt{\Delta \nu T}}
\end{equation}

where $n_{bin}$ is the number of phase bins in the average pulse profile, $\Delta \nu$ is the observing bandwidth, $T$ is total integration time of typically 290\,s (thus the integration time per phase bin of the pulse profile is $\Delta t = T / n_{bin}$), and $\text{SEFD}_I$ is Stokes I polarisation calculated from $\text{SEFD}_X$ and $\text{SEFD}_Y$ according to equation~\ref{eq_sefd_i}.

\item Calculate calibration coefficient $z = \frac{\sigma_s}{\sigma_o}$, and use it to calculate calibrated flux density in each phase bin $i$ according to: $f_c^i = z \cdot f_u^i$, where $f_u^i$ and $f_c^i$ are respectively uncalibrated and calibrated flux densities in the phase bin $i$.

\item Calculate calibrated mean flux density as:

\begin{equation}
m = \frac{F}{P} = \frac{\text{SEFD}_I \times \sum_{i=0}^{n_{bin}} f_u^{i}}{\sigma_o \sqrt{\Delta \nu \cdot T \cdot n_{bin}}},
\end{equation}

where $F$ is fluence and $P$ pulsar period.

\item Calculate calibrated peak flux $f_p$ as: 

\begin{equation}
f_p = \max\limits_{1\leq i\leq n_{bin}} [ f_c^{i} ]
\end{equation}

\end{enumerate}

\end{appendix}

\bibliographystyle{pasa-mnras}
\bibliography{mwa_frb}

\begin{thebibliography}{}
\makeatletter
\relax
\def\mn@urlcharsother{\let\do\@makeother \do\$\do\&\do\#\do\^\do\_\do\%\do\~}
\definecolor{darkblue}{rgb}{0,0,0.597656}
\def\mndoi{\begingroup\mn@urlcharsother \@ifnextchar [ {\mndoi@} {\mndoi@[]}}
\def\mndoi@[#1]#2{\def\@tempa{#1}\ifx\@tempa\@empty \href {http://dx.doi.org/#2} {\textcolor{darkblue}{doi:#2}}\else \href {http://dx.doi.org/#2} {\textcolor{darkblue}{#1}}\fi \endgroup}
\def\mn@eprint#1#2{\mn@eprint@#1:#2::\@nil}
\def\mn@eprint@arXiv#1{\href {http://arxiv.org/abs/#1} {{\tt arXiv:#1}}}
\def\mn@eprint@dblp#1{\href {http://dblp.uni-trier.de/rec/bibtex/#1.xml} {dblp:#1}}
\def\mn@eprint@#1:#2:#3:#4\@nil{\def\@tempa {#1}\def\@tempb {#2}\def\@tempc {#3}\ifx \@tempc \@empty \let \@tempc \@tempb \let \@tempb \@tempa \fi \ifx \@tempb \@empty \def\@tempb {arXiv}\fi \@ifundefined {mn@eprint@\@tempb}{\@tempb:\@tempc}{\expandafter \expandafter \csname mn@eprint@\@tempb\endcsname \expandafter{\@tempc}}}

\bibitem[\protect\citeauthoryear{{Abbott} et~al.,}{{Abbott} et~al.}{2017}]{2017ApJ...848L..12A}
{Abbott} B.~P.,  et~al., 2017, \mndoi [ApJL] {10.3847/2041-8213/aa91c9}, \href {https://ui.adsabs.harvard.edu/abs/2017ApJ...848L..12A} {848, L12}

\bibitem[\protect\citeauthoryear{{Abbott} et~al.,}{{Abbott} et~al.}{2018}]{2018LRR....21....3A}
{Abbott} B.~P.,  et~al., 2018, \mndoi [Living Reviews in Relativity] {10.1007/s41114-018-0012-9}, \href {https://ui.adsabs.harvard.edu/abs/2018LRR....21....3A} {21, 3}

\bibitem[\protect\citeauthoryear{{Andersen} et~al.,}{{Andersen} et~al.}{2023}]{2023ApJ...947...83A}
{Andersen} B.~C.,  et~al., 2023, \mndoi [\apj] {10.3847/1538-4357/acc6c1}, \href {https://ui.adsabs.harvard.edu/abs/2023ApJ...947...83A} {947, 83}

\bibitem[\protect\citeauthoryear{{Anderson} et~al.,}{{Anderson} et~al.}{2021}]{2021PASA...38...26A}
{Anderson} G.~E.,  et~al., 2021, \mndoi [\pasa] {10.1017/pasa.2021.15}, \href {https://ui.adsabs.harvard.edu/abs/2021PASA...38...26A} {38, e026}

\bibitem[\protect\citeauthoryear{{Bailes} et~al.,}{{Bailes} et~al.}{2017}]{2017PASA...34...45B}
{Bailes} M.,  et~al., 2017, \mndoi [\pasa] {10.1017/pasa.2017.39}, \href {https://ui.adsabs.harvard.edu/abs/2017PASA...34...45B} {34, e045}

\bibitem[\protect\citeauthoryear{{Bannister}, {Zackay}, {Qiu}, {James}  \& {Shannon}}{{Bannister} et~al.}{2019a}]{2019ascl.soft06003B}
{Bannister} K.,  {Zackay} B.,  {Qiu} H.,  {James} C.,   {Shannon} R.,  2019a, {FREDDA: A fast, real-time engine for de-dispersing amplitudes}, Astrophysics Source Code Library, record ascl:1906.003 (\mn@eprint {ascl} {1906.003})

\bibitem[\protect\citeauthoryear{{Bannister} et~al.,}{{Bannister} et~al.}{2019b}]{2019Sci...365..565B}
{Bannister} K.~W.,  et~al., 2019b, \mndoi [Science] {10.1126/science.aaw5903}, \href {https://ui.adsabs.harvard.edu/abs/2019Sci...365..565B} {365, 565}

\bibitem[\protect\citeauthoryear{{Bell} et~al.,}{{Bell} et~al.}{2016}]{2016MNRAS.461..908B}
{Bell} M.~E.,  et~al., 2016, \mndoi [\mnras] {10.1093/mnras/stw1293}, \href {https://ui.adsabs.harvard.edu/abs/2016MNRAS.461..908B} {461, 908}

\bibitem[\protect\citeauthoryear{{Bhandari} et~al.,}{{Bhandari} et~al.}{2018}]{2018MNRAS.475.1427B}
{Bhandari} S.,  et~al., 2018, \mndoi [\mnras] {10.1093/mnras/stx3074}, \href {https://ui.adsabs.harvard.edu/abs/2018MNRAS.475.1427B} {475, 1427}

\bibitem[\protect\citeauthoryear{{Bhat} et~al.,}{{Bhat} et~al.}{2023a}]{2023PASA...40...20B}
{Bhat} N.~D.~R.,  et~al., 2023a, \mndoi [\pasa] {10.1017/pasa.2023.18}, \href {https://ui.adsabs.harvard.edu/abs/2023PASA...40...20B} {40, e020}

\bibitem[\protect\citeauthoryear{{Bhat} et~al.,}{{Bhat} et~al.}{2023b}]{2023PASA...40...21B}
{Bhat} N.~D.~R.,  et~al., 2023b, \mndoi [\pasa] {10.1017/pasa.2023.17}, \href {https://ui.adsabs.harvard.edu/abs/2023PASA...40...21B} {40, e021}

\bibitem[\protect\citeauthoryear{{Bochenek}, {Ravi}, {Belov}, {Hallinan}, {Kocz}, {Kulkarni}  \& {McKenna}}{{Bochenek} et~al.}{2020}]{2020Natur.587...59B}
{Bochenek} C.~D.,  {Ravi} V.,  {Belov} K.~V.,  {Hallinan} G.,  {Kocz} J.,  {Kulkarni} S.~R.,   {McKenna} D.~L.,  2020, \mndoi [\nat] {10.1038/s41586-020-2872-x}, \href {https://ui.adsabs.harvard.edu/abs/2020Natur.587...59B} {587, 59}

\bibitem[\protect\citeauthoryear{{CHIME/FRB Collaboration} et~al.,}{{CHIME/FRB Collaboration} et~al.}{2020a}]{2020Natur.582..351C}
{CHIME/FRB Collaboration} et~al., 2020a, \mndoi [\nat] {10.1038/s41586-020-2398-2}, \href {https://ui.adsabs.harvard.edu/abs/2020Natur.582..351C} {582, 351}

\bibitem[\protect\citeauthoryear{{CHIME/FRB Collaboration} et~al.,}{{CHIME/FRB Collaboration} et~al.}{2020b}]{2020Natur.587...54C}
{CHIME/FRB Collaboration} et~al., 2020b, \mndoi [\nat] {10.1038/s41586-020-2863-y}, \href {https://ui.adsabs.harvard.edu/abs/2020Natur.587...54C} {587, 54}

\bibitem[\protect\citeauthoryear{{CHIME/FRB Collaboration} et~al.,}{{CHIME/FRB Collaboration} et~al.}{2021}]{2021ApJS..257...59C}
{CHIME/FRB Collaboration} et~al., 2021, \mndoi [\apjs] {10.3847/1538-4365/ac33ab}, \href {https://ui.adsabs.harvard.edu/abs/2021ApJS..257...59C} {257, 59}

\bibitem[\protect\citeauthoryear{{Caleb} et~al.,}{{Caleb} et~al.}{2017}]{2017MNRAS.468.3746C}
{Caleb} M.,  et~al., 2017, \mndoi [\mnras] {10.1093/mnras/stx638}, \href {https://ui.adsabs.harvard.edu/abs/2017MNRAS.468.3746C} {468, 3746}

\bibitem[\protect\citeauthoryear{{Chatterjee} et~al.,}{{Chatterjee} et~al.}{2017}]{2017Natur.541...58C}
{Chatterjee} S.,  et~al., 2017, \mndoi [\nat] {10.1038/nature20797}, \href {https://ui.adsabs.harvard.edu/abs/2017Natur.541...58C} {541, 58}

\bibitem[\protect\citeauthoryear{{Chu}, {Howell}, {Rowlinson}, {Gao}, {Zhang}, {Tingay}, {Bo{\"e}r}  \& {Wen}}{{Chu} et~al.}{2016}]{2016MNRAS.459..121C}
{Chu} Q.,  {Howell} E.~J.,  {Rowlinson} A.,  {Gao} H.,  {Zhang} B.,  {Tingay} S.~J.,  {Bo{\"e}r} M.,   {Wen} L.,  2016, \mndoi [\mnras] {10.1093/mnras/stw576}, \href {https://ui.adsabs.harvard.edu/abs/2016MNRAS.459..121C} {459, 121}

\bibitem[\protect\citeauthoryear{{Coenen} et~al.,}{{Coenen} et~al.}{2014}]{Coenen_et_al_2014}
{Coenen} T.,  et~al., 2014, \mndoi [\aap] {10.1051/0004-6361/201424495}, \href {http://adsabs.harvard.edu/abs/2014A%26A...570A..60C} {570, A60}

\bibitem[\protect\citeauthoryear{{Connor}, {Pen}  \& {Oppermann}}{{Connor} et~al.}{2016}]{2016MNRAS.458L..89C}
{Connor} L.,  {Pen} U.-L.,   {Oppermann} N.,  2016, \mndoi [\mnras] {10.1093/mnrasl/slw026}, \href {https://ui.adsabs.harvard.edu/abs/2016MNRAS.458L..89C} {458, L89}

\bibitem[\protect\citeauthoryear{{Cook}, {Seymour}  \& {Sokolowski}}{{Cook} et~al.}{2021}]{2021PASA...38...63C}
{Cook} J.~H.,  {Seymour} N.,   {Sokolowski} M.,  2021, \mndoi [\pasa] {10.1017/pasa.2021.55}, \href {https://ui.adsabs.harvard.edu/abs/2021PASA...38...63C} {38, e063}

\bibitem[\protect\citeauthoryear{{Cordes}}{{Cordes}}{2008}]{2008ASPC..395..225C}
{Cordes} J.~M.,  2008, in {Bridle} A.~H.,  {Condon} J.~J.,   {Hunt} G.~C.,  eds,  Astronomical Society of the Pacific Conference Series Vol. 395, Frontiers of Astrophysics: A Celebration of NRAO's 50th Anniversary. p.~225

\bibitem[\protect\citeauthoryear{{Cordes} \& {Chatterjee}}{{Cordes} \& {Chatterjee}}{2019}]{2019ARA&A..57..417C}
{Cordes} J.~M.,  {Chatterjee} S.,  2019, \mndoi [\araa] {10.1146/annurev-astro-091918-104501}, \href {https://ui.adsabs.harvard.edu/abs/2019ARA&A..57..417C} {57, 417}

\bibitem[\protect\citeauthoryear{{Cordes} \& {Wasserman}}{{Cordes} \& {Wasserman}}{2016}]{2016MNRAS.457..232C}
{Cordes} J.~M.,  {Wasserman} I.,  2016, \mndoi [\mnras] {10.1093/mnras/stv2948}, \href {https://ui.adsabs.harvard.edu/abs/2016MNRAS.457..232C} {457, 232}

\bibitem[\protect\citeauthoryear{{Cordes}, {Lazio}  \& {McLaughlin}}{{Cordes} et~al.}{2004}]{2004NewAR..48.1459C}
{Cordes} J.~M.,  {Lazio} T. J.~W.,   {McLaughlin} M.~A.,  2004, \mndoi [New Astronomy Reviews] {10.1016/j.newar.2004.09.038}, \href {https://ui.adsabs.harvard.edu/abs/2004NewAR..48.1459C} {48, 1459}

\bibitem[\protect\citeauthoryear{{Dewdney}, {Hall}, {Schilizzi}  \& {Lazio}}{{Dewdney} et~al.}{2009}]{2009IEEEP..97.1482D}
{Dewdney} P.~E.,  {Hall} P.~J.,  {Schilizzi} R.~T.,   {Lazio} T.~J.~L.~W.,  2009, \mndoi [IEEE Proceedings] {10.1109/JPROC.2009.2021005}, \href {https://ui.adsabs.harvard.edu/abs/2009IEEEP..97.1482D} {97, 1482}

\bibitem[\protect\citeauthoryear{{Falcke} \& {Rezzolla}}{{Falcke} \& {Rezzolla}}{2014}]{2014A&A...562A.137F}
{Falcke} H.,  {Rezzolla} L.,  2014, \mndoi [\aap] {10.1051/0004-6361/201321996}, \href {https://ui.adsabs.harvard.edu/abs/2014A&A...562A.137F} {562, A137}

\bibitem[\protect\citeauthoryear{{Farah} et~al.,}{{Farah} et~al.}{2019}]{2019MNRAS.488.2989F}
{Farah} W.,  et~al., 2019, \mndoi [\mnras] {10.1093/mnras/stz1748}, \href {https://ui.adsabs.harvard.edu/abs/2019MNRAS.488.2989F} {488, 2989}

\bibitem[\protect\citeauthoryear{{Fong}, {Berger}, {Margutti}  \& {Zauderer}}{{Fong} et~al.}{2015}]{2015ApJ...815..102F}
{Fong} W.,  {Berger} E.,  {Margutti} R.,   {Zauderer} B.~A.,  2015, \mndoi [\apj] {10.1088/0004-637X/815/2/102}, \href {https://ui.adsabs.harvard.edu/abs/2015ApJ...815..102F} {815, 102}

\bibitem[\protect\citeauthoryear{{Hancock} et~al.,}{{Hancock} et~al.}{2019}]{2019PASA...36...46H}
{Hancock} P.~J.,  et~al., 2019, \mndoi [\pasa] {10.1017/pasa.2019.40}, \href {https://ui.adsabs.harvard.edu/abs/2019PASA...36...46H} {36, e046}

\bibitem[\protect\citeauthoryear{{Hessels}, {Stappers}, {van Leeuwen}  \& {LOFAR}}{{Hessels} et~al.}{2009}]{2009ASPC..407..318H}
{Hessels} J.~W.~T.,  {Stappers} B.~W.,  {van Leeuwen} J.,   {LOFAR} 2009, in {Saikia} D.~J.,  {Green} D.~A.,  {Gupta} Y.,   {Venturi} T.,  eds,  Astronomical Society of the Pacific Conference Series Vol. 407, The Low-Frequency Radio Universe. p.~318 (\mn@eprint {arXiv} {0903.1447}), \mndoi{10.48550/arXiv.0903.1447}

\bibitem[\protect\citeauthoryear{{Hurley-Walker} et~al.,}{{Hurley-Walker} et~al.}{2022}]{2022Natur.601..526H}
{Hurley-Walker} N.,  et~al., 2022, \mndoi [\nat] {10.1038/s41586-021-04272-x}, \href {https://ui.adsabs.harvard.edu/abs/2022Natur.601..526H} {601, 526}

\bibitem[\protect\citeauthoryear{{Hurley-Walker} et~al.,}{{Hurley-Walker} et~al.}{2023}]{2023Natur.619..487H}
{Hurley-Walker} N.,  et~al., 2023, \mndoi [\nat] {10.1038/s41586-023-06202-5}, \href {https://ui.adsabs.harvard.edu/abs/2023Natur.619..487H} {619, 487}

\bibitem[\protect\citeauthoryear{{Ioka} \& {Zhang}}{{Ioka} \& {Zhang}}{2020}]{2020ApJ...893L..26I}
{Ioka} K.,  {Zhang} B.,  2020, \mndoi [\apjl] {10.3847/2041-8213/ab83fb}, \href {https://ui.adsabs.harvard.edu/abs/2020ApJ...893L..26I} {893, L26}

\bibitem[\protect\citeauthoryear{{James}, {Anderson}, {Wen}, {Bosveld}, {Chu}, {Kovalam}, {Slaven-Blair}  \& {Williams}}{{James} et~al.}{2019}]{2019MNRAS.489L..75J}
{James} C.~W.,  {Anderson} G.~E.,  {Wen} L.,  {Bosveld} J.,  {Chu} Q.,  {Kovalam} M.,  {Slaven-Blair} T.~J.,   {Williams} A.,  2019, \mndoi [\mnras] {10.1093/mnrasl/slz129}, \href {https://ui.adsabs.harvard.edu/abs/2019MNRAS.489L..75J} {489, L75}

\bibitem[\protect\citeauthoryear{{James}, {Prochaska}, {Macquart}, {North-Hickey}, {Bannister}  \& {Dunning}}{{James} et~al.}{2022}]{2022MNRAS.509.4775J}
{James} C.~W.,  {Prochaska} J.~X.,  {Macquart} J.~P.,  {North-Hickey} F.~O.,  {Bannister} K.~W.,   {Dunning} A.,  2022, \mndoi [\mnras] {10.1093/mnras/stab3051}, \href {https://ui.adsabs.harvard.edu/abs/2022MNRAS.509.4775J} {509, 4775}

\bibitem[\protect\citeauthoryear{{Karastergiou} et~al.,}{{Karastergiou} et~al.}{2015}]{2015MNRAS.452.1254K}
{Karastergiou} A.,  et~al., 2015, \mndoi [\mnras] {10.1093/mnras/stv1306}, \href {http://adsabs.harvard.edu/abs/2015MNRAS.452.1254K} {452, 1254}

\bibitem[\protect\citeauthoryear{{Keane} et~al.,}{{Keane} et~al.}{2016}]{2016Natur.530..453K}
{Keane} E.~F.,  et~al., 2016, \mndoi [\nat] {10.1038/nature17140}, \href {http://adsabs.harvard.edu/abs/2016Natur.530..453K} {530, 453}

\bibitem[\protect\citeauthoryear{{Kirsten} et~al.,}{{Kirsten} et~al.}{2022}]{2022Natur.602..585K}
{Kirsten} F.,  et~al., 2022, \mndoi [\nat] {10.1038/s41586-021-04354-w}, \href {https://ui.adsabs.harvard.edu/abs/2022Natur.602..585K} {602, 585}

\bibitem[\protect\citeauthoryear{{Kuiack}, {Wijers}, {Rowlinson}, {Shulevski}, {Huizinga}, {Molenaar}  \& {Prasad}}{{Kuiack} et~al.}{2020}]{2020MNRAS.497..846K}
{Kuiack} M.,  {Wijers} R. A.~M.~J.,  {Rowlinson} A.,  {Shulevski} A.,  {Huizinga} F.,  {Molenaar} G.,   {Prasad} P.,  2020, \mndoi [\mnras] {10.1093/mnras/staa1996}, \href {https://ui.adsabs.harvard.edu/abs/2020MNRAS.497..846K} {497, 846}

\bibitem[\protect\citeauthoryear{{Kumar}, {Lu}  \& {Bhattacharya}}{{Kumar} et~al.}{2017}]{2017MNRAS.468.2726K}
{Kumar} P.,  {Lu} W.,   {Bhattacharya} M.,  2017, \mndoi [\mnras] {10.1093/mnras/stx665}, \href {https://ui.adsabs.harvard.edu/abs/2017MNRAS.468.2726K} {468, 2726}

\bibitem[\protect\citeauthoryear{{Large}, {Vaughan}  \& {Mills}}{{Large} et~al.}{1968}]{1968Natur.220..340L}
{Large} M.~I.,  {Vaughan} A.~E.,   {Mills} B.~Y.,  1968, \mndoi [\nat] {10.1038/220340a0}, \href {https://ui.adsabs.harvard.edu/abs/1968Natur.220..340L} {220, 340}

\bibitem[\protect\citeauthoryear{{Lee}, {Bhat}, {Sokolowski}, {Swainston}, {Ung}, {Magro}  \& {Chiello}}{{Lee} et~al.}{2022}]{2022PASA...39...42L}
{Lee} C.~P.,  {Bhat} N.~D.~R.,  {Sokolowski} M.,  {Swainston} N.~A.,  {Ung} D.,  {Magro} A.,   {Chiello} R.,  2022, \mndoi [\pasa] {10.1017/pasa.2022.40}, \href {https://ui.adsabs.harvard.edu/abs/2022PASA...39...42L} {39, e042}

\bibitem[\protect\citeauthoryear{{Lin} et~al.,}{{Lin} et~al.}{2023a}]{2023arXiv230705261L}
{Lin} H.-H.,  et~al., 2023a, \mndoi [arXiv e-prints] {10.48550/arXiv.2307.05261}, \href {https://ui.adsabs.harvard.edu/abs/2023arXiv230705261L} {p. arXiv:2307.05261}

\bibitem[\protect\citeauthoryear{{Lin} et~al.,}{{Lin} et~al.}{2023b}]{2023arXiv230705262L}
{Lin} H.-H.,  et~al., 2023b, \mndoi [arXiv e-prints] {10.48550/arXiv.2307.05262}, \href {https://ui.adsabs.harvard.edu/abs/2023arXiv230705262L} {p. arXiv:2307.05262}

\bibitem[\protect\citeauthoryear{{Liu}}{{Liu}}{2018}]{2018Ap&SS.363..242L}
{Liu} X.,  2018, \mndoi [\apss] {10.1007/s10509-018-3462-3}, \href {https://ui.adsabs.harvard.edu/abs/2018Ap&SS.363..242L} {363, 242}

\bibitem[\protect\citeauthoryear{{Lorimer}, {Bailes}, {McLaughlin}, {Narkevic}  \& {Crawford}}{{Lorimer} et~al.}{2007}]{2007Sci...318..777L}
{Lorimer} D.~R.,  {Bailes} M.,  {McLaughlin} M.~A.,  {Narkevic} D.~J.,   {Crawford} F.,  2007, \mndoi [Science] {10.1126/science.1147532}, \href {https://ui.adsabs.harvard.edu/abs/2007Sci...318..777L} {318, 777}

\bibitem[\protect\citeauthoryear{{Lyutikov}}{{Lyutikov}}{2021}]{2021ApJ...922..166L}
{Lyutikov} M.,  2021, \mndoi [\apj] {10.3847/1538-4357/ac1b32}, \href {https://ui.adsabs.harvard.edu/abs/2021ApJ...922..166L} {922, 166}

\bibitem[\protect\citeauthoryear{{Lyutikov}, {Barkov}  \& {Giannios}}{{Lyutikov} et~al.}{2020}]{2020ApJ...893L..39L}
{Lyutikov} M.,  {Barkov} M.~V.,   {Giannios} D.,  2020, \mndoi [\apjl] {10.3847/2041-8213/ab87a4}, \href {https://ui.adsabs.harvard.edu/abs/2020ApJ...893L..39L} {893, L39}

\bibitem[\protect\citeauthoryear{{Maan} \& {van Leeuwen}}{{Maan} \& {van Leeuwen}}{2017}]{2017ursi.confE...2M}
{Maan} Y.,  {van Leeuwen} J.,  2017, in 2017 XXXIInd URSI GASS. p.~2 (\mn@eprint {arXiv} {1709.06104}), \mndoi{10.23919/URSIGASS.2017.8105320}

\bibitem[\protect\citeauthoryear{{Macario} et~al.,}{{Macario} et~al.}{2022}]{2022JATIS...8a1014M}
{Macario} G.,  et~al., 2022, \mndoi [SPIE JATIS] {10.1117/1.JATIS.8.1.011014}, \href {https://ui.adsabs.harvard.edu/abs/2022JATIS...8a1014M} {8, 011014}

\bibitem[\protect\citeauthoryear{{Macquart} et~al.,}{{Macquart} et~al.}{2010}]{Macquartetal2010}
{Macquart} J.-P.,  et~al., 2010, \mndoi [\pasa] {10.1071/AS09082}, \href {http://adsabs.harvard.edu/abs/2010PASA...27..272M} {27, 272}

\bibitem[\protect\citeauthoryear{{Macquart} et~al.,}{{Macquart} et~al.}{2020}]{2020Natur.581..391M}
{Macquart} J.~P.,  et~al., 2020, \mndoi [\nat] {10.1038/s41586-020-2300-2}, \href {https://ui.adsabs.harvard.edu/abs/2020Natur.581..391M} {581, 391}

\bibitem[\protect\citeauthoryear{{Manchester}, {Hobbs}, {Teoh}  \& {Hobbs}}{{Manchester} et~al.}{2005}]{2005AJ....129.1993M}
{Manchester} R.~N.,  {Hobbs} G.~B.,  {Teoh} A.,   {Hobbs} M.,  2005, \mndoi [\aj] {10.1086/428488}, \href {https://ui.adsabs.harvard.edu/abs/2005AJ....129.1993M} {129, 1993}

\bibitem[\protect\citeauthoryear{{Marcote} et~al.,}{{Marcote} et~al.}{2020}]{2020Natur.577..190M}
{Marcote} B.,  et~al., 2020, \mndoi [\nat] {10.1038/s41586-019-1866-z}, \href {https://ui.adsabs.harvard.edu/abs/2020Natur.577..190M} {577, 190}

\bibitem[\protect\citeauthoryear{{Margalit}, {Metzger}, {Berger}, {Nicholl}, {Eftekhari}  \& {Margutti}}{{Margalit} et~al.}{2018}]{2018MNRAS.481.2407M}
{Margalit} B.,  {Metzger} B.~D.,  {Berger} E.,  {Nicholl} M.,  {Eftekhari} T.,   {Margutti} R.,  2018, \mndoi [\mnras] {10.1093/mnras/sty2417}, \href {https://ui.adsabs.harvard.edu/abs/2018MNRAS.481.2407M} {481, 2407}

\bibitem[\protect\citeauthoryear{{McSweeney} et~al.,}{{McSweeney} et~al.}{2020}]{2020PASA...37...34M}
{McSweeney} S.~J.,  et~al., 2020, \mndoi [\pasa] {10.1017/pasa.2020.24}, \href {https://ui.adsabs.harvard.edu/abs/2020PASA...37...34M} {37, e034}

\bibitem[\protect\citeauthoryear{{Metzger}}{{Metzger}}{2017}]{2017LRR....20....3M}
{Metzger} B.~D.,  2017, \mndoi [Living Reviews in Relativity] {10.1007/s41114-017-0006-z}, \href {https://ui.adsabs.harvard.edu/abs/2017LRR....20....3M} {20, 3}

\bibitem[\protect\citeauthoryear{{Metzger}, {Berger}  \& {Margalit}}{{Metzger} et~al.}{2017}]{2017ApJ...841...14M}
{Metzger} B.~D.,  {Berger} E.,   {Margalit} B.,  2017, \mndoi [\apj] {10.3847/1538-4357/aa633d}, \href {https://ui.adsabs.harvard.edu/abs/2017ApJ...841...14M} {841, 14}

\bibitem[\protect\citeauthoryear{{Metzger}, {Margalit}  \& {Sironi}}{{Metzger} et~al.}{2019}]{2019MNRAS.485.4091M}
{Metzger} B.~D.,  {Margalit} B.,   {Sironi} L.,  2019, \mndoi [\mnras] {10.1093/mnras/stz700}, \href {https://ui.adsabs.harvard.edu/abs/2019MNRAS.485.4091M} {485, 4091}

\bibitem[\protect\citeauthoryear{{Moroianu}, {Wen}, {James}, {Ai}, {Kovalam}, {Panther}  \& {Zhang}}{{Moroianu} et~al.}{2023}]{2023NatAs...7..579M}
{Moroianu} A.,  {Wen} L.,  {James} C.~W.,  {Ai} S.,  {Kovalam} M.,  {Panther} F.~H.,   {Zhang} B.,  2023, \mndoi [Nature Astronomy] {10.1038/s41550-023-01917-x}, \href {https://ui.adsabs.harvard.edu/abs/2023NatAs...7..579M} {7, 579}

\bibitem[\protect\citeauthoryear{{Morrison} et~al.,}{{Morrison} et~al.}{2023}]{2023PASA...40...19M}
{Morrison} I.~S.,  et~al., 2023, \mndoi [\pasa] {10.1017/pasa.2023.15}, \href {https://ui.adsabs.harvard.edu/abs/2023PASA...40...19M} {40, e019}

\bibitem[\protect\citeauthoryear{{Ord} et~al.,}{{Ord} et~al.}{2015}]{2015PASA...32....6O}
{Ord} S.~M.,  et~al., 2015, \mndoi [\pasa] {10.1017/pasa.2015.5}, \href {https://ui.adsabs.harvard.edu/abs/2015PASA...32....6O} {32, e006}

\bibitem[\protect\citeauthoryear{{Ord}, {Tremblay}, {McSweeney}, {Bhat}, {Sobey}, {Mitchell}, {Hancock}  \& {Kirsten}}{{Ord} et~al.}{2019}]{2019PASA...36...30O}
{Ord} S.~M.,  {Tremblay} S.~E.,  {McSweeney} S.~J.,  {Bhat} N.~D.~R.,  {Sobey} C.,  {Mitchell} D.~A.,  {Hancock} P.~J.,   {Kirsten} F.,  2019, \mndoi [\pasa] {10.1017/pasa.2019.17}, \href {https://ui.adsabs.harvard.edu/abs/2019PASA...36...30O} {36, e030}

\bibitem[\protect\citeauthoryear{{Panther} et~al.,}{{Panther} et~al.}{2023}]{2023MNRAS.519.2235P}
{Panther} F.~H.,  et~al., 2023, \mndoi [\mnras] {10.1093/mnras/stac3597}, \href {https://ui.adsabs.harvard.edu/abs/2023MNRAS.519.2235P} {519, 2235}

\bibitem[\protect\citeauthoryear{{Parent} et~al.,}{{Parent} et~al.}{2020}]{2020ApJ...904...92P}
{Parent} E.,  et~al., 2020, \mndoi [\apj] {10.3847/1538-4357/abbdf6}, \href {https://ui.adsabs.harvard.edu/abs/2020ApJ...904...92P} {904, 92}

\bibitem[\protect\citeauthoryear{{Pastor-Marazuela} et~al.,}{{Pastor-Marazuela} et~al.}{2021}]{2021Natur.596..505P}
{Pastor-Marazuela} I.,  et~al., 2021, \mndoi [\nat] {10.1038/s41586-021-03724-8}, \href {https://ui.adsabs.harvard.edu/abs/2021Natur.596..505P} {596, 505}

\bibitem[\protect\citeauthoryear{{Petroff}, {Hessels}  \& {Lorimer}}{{Petroff} et~al.}{2019}]{2019A&ARv..27....4P}
{Petroff} E.,  {Hessels} J.~W.~T.,   {Lorimer} D.~R.,  2019, \mndoi [\aapr] {10.1007/s00159-019-0116-6}, \href {https://ui.adsabs.harvard.edu/abs/2019A&ARv..27....4P} {27, 4}

\bibitem[\protect\citeauthoryear{{Petroff}, {Hessels}  \& {Lorimer}}{{Petroff} et~al.}{2022}]{2022A&ARv..30....2P}
{Petroff} E.,  {Hessels} J.~W.~T.,   {Lorimer} D.~R.,  2022, \mndoi [\aapr] {10.1007/s00159-022-00139-w}, \href {https://ui.adsabs.harvard.edu/abs/2022A&ARv..30....2P} {30, 2}

\bibitem[\protect\citeauthoryear{{Pilia}}{{Pilia}}{2021}]{2021Univ....8....9P}
{Pilia} M.,  2021, \mndoi [Universe] {10.3390/universe8010009}, \href {https://ui.adsabs.harvard.edu/abs/2021Univ....8....9P} {8, 9}

\bibitem[\protect\citeauthoryear{{Pilkington}, {Hewish}, {Bell}  \& {Cole}}{{Pilkington} et~al.}{1968}]{1968Natur.218..126P}
{Pilkington} J.~D.~H.,  {Hewish} A.,  {Bell} S.~J.,   {Cole} T.~W.,  1968, \mndoi [\nat] {10.1038/218126a0}, \href {https://ui.adsabs.harvard.edu/abs/1968Natur.218..126P} {218, 126}

\bibitem[\protect\citeauthoryear{{Platts}, {Weltman}, {Walters}, {Tendulkar}, {Gordin}  \& {Kandhai}}{{Platts} et~al.}{2019}]{2019PhR...821....1P}
{Platts} E.,  {Weltman} A.,  {Walters} A.,  {Tendulkar} S.~P.,  {Gordin} J.~E.~B.,   {Kandhai} S.,  2019, \mndoi [\physrep] {10.1016/j.physrep.2019.06.003}, \href {https://ui.adsabs.harvard.edu/abs/2019PhR...821....1P} {821, 1}

\bibitem[\protect\citeauthoryear{{Pleunis} et~al.,}{{Pleunis} et~al.}{2021a}]{2021ApJ...911L...3P}
{Pleunis} Z.,  et~al., 2021a, \mndoi [\apjl] {10.3847/2041-8213/abec72}, \href {https://ui.adsabs.harvard.edu/abs/2021ApJ...911L...3P} {911, L3}

\bibitem[\protect\citeauthoryear{{Pleunis} et~al.,}{{Pleunis} et~al.}{2021b}]{2021ApJ...923....1P}
{Pleunis} Z.,  et~al., 2021b, \mndoi [\apj] {10.3847/1538-4357/ac33ac}, \href {https://ui.adsabs.harvard.edu/abs/2021ApJ...923....1P} {923, 1}

\bibitem[\protect\citeauthoryear{{Prochaska} et~al.,}{{Prochaska} et~al.}{2019}]{2019Sci...366..231P}
{Prochaska} J.~X.,  et~al., 2019, \mndoi [Science] {10.1126/science.aay0073}, \href {https://ui.adsabs.harvard.edu/abs/2019Sci...366..231P} {366, 231}

\bibitem[\protect\citeauthoryear{{Rajwade}, {Mickaliger}, {Stappers}, {Bassa}, {Breton}, {Karastergiou}  \& {Keane}}{{Rajwade} et~al.}{2020}]{2020MNRAS.493.4418R}
{Rajwade} K.~M.,  {Mickaliger} M.~B.,  {Stappers} B.~W.,  {Bassa} C.~G.,  {Breton} R.~P.,  {Karastergiou} A.,   {Keane} E.~F.,  2020, \mndoi [\mnras] {10.1093/mnras/staa616}, \href {https://ui.adsabs.harvard.edu/abs/2020MNRAS.493.4418R} {493, 4418}

\bibitem[\protect\citeauthoryear{{Ransom}}{{Ransom}}{2011}]{2011ascl.soft07017R}
{Ransom} S.,  2011, {PRESTO: PulsaR Exploration and Search TOolkit}, Astrophysics Source Code Library, record ascl:1107.017 (\mn@eprint {ascl} {1107.017})

\bibitem[\protect\citeauthoryear{{Ravi} et~al.,}{{Ravi} et~al.}{2019}]{2019Natur.572..352R}
{Ravi} V.,  et~al., 2019, \mndoi [\nat] {10.1038/s41586-019-1389-7}, \href {https://ui.adsabs.harvard.edu/abs/2019Natur.572..352R} {572, 352}

\bibitem[\protect\citeauthoryear{{Ravi} et~al.,}{{Ravi} et~al.}{2023}]{2023ApJ...949L...3R}
{Ravi} V.,  et~al., 2023, \mndoi [\apjl] {10.3847/2041-8213/acc4b6}, \href {https://ui.adsabs.harvard.edu/abs/2023ApJ...949L...3R} {949, L3}

\bibitem[\protect\citeauthoryear{{Rowlinson} \& {Anderson}}{{Rowlinson} \& {Anderson}}{2019}]{2019MNRAS.489.3316R}
{Rowlinson} A.,  {Anderson} G.~E.,  2019, \mndoi [\mnras] {10.1093/mnras/stz2295}, \href {https://ui.adsabs.harvard.edu/abs/2019MNRAS.489.3316R} {489, 3316}

\bibitem[\protect\citeauthoryear{{Rowlinson} et~al.,}{{Rowlinson} et~al.}{2016}]{2016MNRAS.458.3506R}
{Rowlinson} A.,  et~al., 2016, \mndoi [\mnras] {10.1093/mnras/stw451}, \href {http://adsabs.harvard.edu/abs/2016MNRAS.458.3506R} {458, 3506}

\bibitem[\protect\citeauthoryear{{Rowlinson} et~al.,}{{Rowlinson} et~al.}{2023}]{2023arXiv231204237R}
{Rowlinson} A.,  et~al., 2023, \mndoi [arXiv e-prints] {10.48550/arXiv.2312.04237}, \href {https://ui.adsabs.harvard.edu/abs/2023arXiv231204237R} {p. arXiv:2312.04237}

\bibitem[\protect\citeauthoryear{Ryder et~al.,}{Ryder et~al.}{2023}]{doi:10.1126/science.adf2678}
Ryder S.~D.,  et~al., 2023, \mndoi [Science] {10.1126/science.adf2678}, 382, 294

\bibitem[\protect\citeauthoryear{{Sanghavi} et~al.,}{{Sanghavi} et~al.}{2023}]{2023arXiv230410534S}
{Sanghavi} P.,  et~al., 2023, \mndoi [arXiv e-prints] {10.48550/arXiv.2304.10534}, \href {https://ui.adsabs.harvard.edu/abs/2023arXiv230410534S} {p. arXiv:2304.10534}

\bibitem[\protect\citeauthoryear{{Shannon} et~al.,}{{Shannon} et~al.}{2018}]{2018Natur.562..386S}
{Shannon} R.~M.,  et~al., 2018, \mndoi [\nat] {10.1038/s41586-018-0588-y}, \href {https://ui.adsabs.harvard.edu/abs/2018Natur.562..386S} {562, 386}

\bibitem[\protect\citeauthoryear{{Sokolowski} et~al.,}{{Sokolowski} et~al.}{2017}]{2017PASA...34...62S}
{Sokolowski} M.,  et~al., 2017, \mndoi [\pasa] {10.1017/pasa.2017.54}, \href {https://ui.adsabs.harvard.edu/abs/2017PASA...34...62S} {34, e062}

\bibitem[\protect\citeauthoryear{{Sokolowski} et~al.,}{{Sokolowski} et~al.}{2018}]{2018ApJ...867L..12S}
{Sokolowski} M.,  et~al., 2018, \mndoi [ApJL] {10.3847/2041-8213/aae58d}, \href {https://ui.adsabs.harvard.edu/abs/2018ApJ...867L..12S} {867, L12}

\bibitem[\protect\citeauthoryear{{Sokolowski} et~al.,}{{Sokolowski} et~al.}{2021}]{2021PASA...38...23S}
{Sokolowski} M.,  et~al., 2021, \mndoi [\pasa] {10.1017/pasa.2021.16}, \href {https://ui.adsabs.harvard.edu/abs/2021PASA...38...23S} {38, e023}

\bibitem[\protect\citeauthoryear{{Sokolowski}, {Price}  \& {Wayth}}{{Sokolowski} et~al.}{2022a}]{2022aapr.confE...1S}
{Sokolowski} M.,  {Price} D.~C.,   {Wayth} Randall B.,  2022a, in 2022 3rd URSI Atlantic and Asia Pacific Radio Science Meeting (AT-AP-RASC. p.~1, \mndoi{10.23919/AT-AP-RASC54737.2022.9814380}

\bibitem[\protect\citeauthoryear{{Sokolowski} et~al.,}{{Sokolowski} et~al.}{2022b}]{2022PASA...39...15S}
{Sokolowski} M.,  et~al., 2022b, \mndoi [\pasa] {10.1017/pasa.2021.63}, \href {https://ui.adsabs.harvard.edu/abs/2022PASA...39...15S} {39, e015}

\bibitem[\protect\citeauthoryear{{Spitler} et~al.,}{{Spitler} et~al.}{2014}]{2014ApJ...790..101S}
{Spitler} L.~G.,  et~al., 2014, \mndoi [\apj] {10.1088/0004-637X/790/2/101}, \href {https://ui.adsabs.harvard.edu/abs/2014ApJ...790..101S} {790, 101}

\bibitem[\protect\citeauthoryear{{Sutinjo}, {Sokolowski}, {Kovaleva}, {Ung}, {Broderick}, {Wayth}, {Davidson}  \& {Tingay}}{{Sutinjo} et~al.}{2021}]{2021A&A...646A.143S}
{Sutinjo} A.~T.,  {Sokolowski} M.,  {Kovaleva} M.,  {Ung} D.~C.~X.,  {Broderick} J.~W.,  {Wayth} R.~B.,  {Davidson} D.~B.,   {Tingay} S.~J.,  2021, \mndoi [\aap] {10.1051/0004-6361/202039445}, \href {https://ui.adsabs.harvard.edu/abs/2021A&A...646A.143S} {646, A143}

\bibitem[\protect\citeauthoryear{{Sutinjo}, {Ung}, {Sokolowski}, {Kovaleva}  \& {McSweeney}}{{Sutinjo} et~al.}{2022}]{2022A&A...660A.134S}
{Sutinjo} A.~T.,  {Ung} D.~C.~X.,  {Sokolowski} M.,  {Kovaleva} M.,   {McSweeney} S.,  2022, \mndoi [\aap] {10.1051/0004-6361/202142759}, \href {https://ui.adsabs.harvard.edu/abs/2022A&A...660A.134S} {660, A134}

\bibitem[\protect\citeauthoryear{{Swainston}, {Bhat}, {Morrison}, {McSweeney}, {Ord}, {Tremblay}  \& {Sokolowski}}{{Swainston} et~al.}{2022}]{2022PASA...39...20S}
{Swainston} N.~A.,  {Bhat} N.~D.~R.,  {Morrison} I.~S.,  {McSweeney} S.~J.,  {Ord} S.~M.,  {Tremblay} S.~E.,   {Sokolowski} M.,  2022, \mndoi [\pasa] {10.1017/pasa.2022.14}, \href {https://ui.adsabs.harvard.edu/abs/2022PASA...39...20S} {39, e020}

\bibitem[\protect\citeauthoryear{{Tendulkar} et~al.,}{{Tendulkar} et~al.}{2017}]{2017ApJ...834L...7T}
{Tendulkar} S.~P.,  et~al., 2017, \mndoi [\apjl] {10.3847/2041-8213/834/2/L7}, \href {https://ui.adsabs.harvard.edu/abs/2017ApJ...834L...7T} {834, L7}

\bibitem[\protect\citeauthoryear{{Thornton} et~al.,}{{Thornton} et~al.}{2013}]{2013Sci...341...53T}
{Thornton} D.,  et~al., 2013, \mndoi [Science] {10.1126/science.1236789}, \href {https://ui.adsabs.harvard.edu/abs/2013Sci...341...53T} {341, 53}

\bibitem[\protect\citeauthoryear{{Tian} et~al.,}{{Tian} et~al.}{2022a}]{2022PASA...39....3T}
{Tian} J.,  et~al., 2022a, \mndoi [\pasa] {10.1017/pasa.2021.58}, \href {https://ui.adsabs.harvard.edu/abs/2022PASA...39....3T} {39, e003}

\bibitem[\protect\citeauthoryear{{Tian} et~al.,}{{Tian} et~al.}{2022b}]{2022MNRAS.514.2756T}
{Tian} J.,  et~al., 2022b, \mndoi [\mnras] {10.1093/mnras/stac1483}, \href {https://ui.adsabs.harvard.edu/abs/2022MNRAS.514.2756T} {514, 2756}

\bibitem[\protect\citeauthoryear{{Tian} et~al.,}{{Tian} et~al.}{2023a}]{2023PASA...40...50T}
{Tian} J.,  et~al., 2023a, \mndoi [\pasa] {10.1017/pasa.2023.49}, \href {https://ui.adsabs.harvard.edu/abs/2023PASA...40...50T} {40, e050}

\bibitem[\protect\citeauthoryear{{Tian} et~al.,}{{Tian} et~al.}{2023b}]{2023MNRAS.518.4278T}
{Tian} J.,  et~al., 2023b, \mndoi [\mnras] {10.1093/mnras/stac3392}, \href {https://ui.adsabs.harvard.edu/abs/2023MNRAS.518.4278T} {518, 4278}

\bibitem[\protect\citeauthoryear{{Tingay} et~al.,}{{Tingay} et~al.}{2013}]{2013PASA...30....7T}
{Tingay} S.~J.,  et~al., 2013, \mndoi [Publications of the Astronomical Society of Australia] {10.1017/pasa.2012.007}, \href {http://adsabs.harvard.edu/abs/2013PASA...30....7T} {30, e007}

\bibitem[\protect\citeauthoryear{{Tingay} et~al.,}{{Tingay} et~al.}{2015}]{2015AJ....150..199T}
{Tingay} S.~J.,  et~al., 2015, \mndoi [\aj] {10.1088/0004-6256/150/6/199}, \href {http://adsabs.harvard.edu/abs/2015AJ....150..199T} {150, 199}

\bibitem[\protect\citeauthoryear{{Tong}, {Wang}  \& {Wang}}{{Tong} et~al.}{2020}]{2020RAA....20..142T}
{Tong} H.,  {Wang} W.,   {Wang} H.-G.,  2020, \mndoi [Research in Astronomy and Astrophysics] {10.1088/1674-4527/20/9/142}, \href {https://ui.adsabs.harvard.edu/abs/2020RAA....20..142T} {20, 142}

\bibitem[\protect\citeauthoryear{{Totani}}{{Totani}}{2013}]{2013PASJ...65L..12T}
{Totani} T.,  2013, \mndoi [\pasj] {10.1093/pasj/65.5.L12}, \href {https://ui.adsabs.harvard.edu/abs/2013PASJ...65L..12T} {65, L12}

\bibitem[\protect\citeauthoryear{{Tremblay} et~al.,}{{Tremblay} et~al.}{2015}]{2015PASA...32....5T}
{Tremblay} S.~E.,  et~al., 2015, \mndoi [\pasa] {10.1017/pasa.2015.6}, \href {https://ui.adsabs.harvard.edu/abs/2015PASA...32....5T} {32, e005}

\bibitem[\protect\citeauthoryear{{Wayth}, {Brisken}, {Deller}, {Majid}, {Thompson}, {Tingay}  \& {Wagstaff}}{{Wayth} et~al.}{2011}]{2011ApJ...735...97W}
{Wayth} R.~B.,  {Brisken} W.~F.,  {Deller} A.~T.,  {Majid} W.~A.,  {Thompson} D.~R.,  {Tingay} S.~J.,   {Wagstaff} K.~L.,  2011, \mndoi [\apj] {10.1088/0004-637X/735/2/97}, \href {https://ui.adsabs.harvard.edu/abs/2011ApJ...735...97W} {735, 97}

\bibitem[\protect\citeauthoryear{{Wayth} et~al.,}{{Wayth} et~al.}{2018}]{2018PASA...35...33W}
{Wayth} R.~B.,  et~al., 2018, \mndoi [\pasa] {10.1017/pasa.2018.37}, \href {https://ui.adsabs.harvard.edu/abs/2018PASA...35...33W} {35, 33}

\bibitem[\protect\citeauthoryear{Wilson, Chow, Harvey-Smith, Indermuehle, Sokolowski  \& Wayth}{Wilson et~al.}{2016}]{7731554}
Wilson C.,  Chow K.,  Harvey-Smith L.,  Indermuehle B.,  Sokolowski M.,   Wayth R.,  2016, in 2016 International Conference on Electromagnetics in Advanced Applications (ICEAA). pp 922--923, \mndoi{10.1109/ICEAA.2016.7731554}

\bibitem[\protect\citeauthoryear{{Xue} et~al.,}{{Xue} et~al.}{2017}]{2017PASA...34...70X}
{Xue} M.,  et~al., 2017, \mndoi [\pasa] {10.1017/pasa.2017.66}, \href {https://ui.adsabs.harvard.edu/abs/2017PASA...34...70X} {34, e070}

\bibitem[\protect\citeauthoryear{{Zanazzi} \& {Lai}}{{Zanazzi} \& {Lai}}{2020}]{2020ApJ...892L..15Z}
{Zanazzi} J.~J.,  {Lai} D.,  2020, \mndoi [\apjl] {10.3847/2041-8213/ab7cdd}, \href {https://ui.adsabs.harvard.edu/abs/2020ApJ...892L..15Z} {892, L15}

\bibitem[\protect\citeauthoryear{{Zhang}}{{Zhang}}{2014}]{2014ApJ...780L..21Z}
{Zhang} B.,  2014, \mndoi [\apjl] {10.1088/2041-8205/780/2/L21}, \href {https://ui.adsabs.harvard.edu/abs/2014ApJ...780L..21Z} {780, L21}

\bibitem[\protect\citeauthoryear{{van Straten}, {Jameson}  \& {Os{\r{A}}owski}}{{van Straten} et~al.}{2021}]{2021ascl.soft10003V}
{van Straten} W.,  {Jameson} A.,   {Os{\r{A}}owski} S.,  2021, {PSRDADA: Distributed Acquisition and Data Analysis for Radio Astronomy}, Astrophysics Source Code Library, record ascl:2110.003 (\mn@eprint {ascl} {2110.003})

\makeatother
\end{thebibliography}

\end{document}